\documentclass{aa}
\usepackage[varg]{txfonts}
\usepackage{graphicx} 
\usepackage{booktabs}
\usepackage{csquotes}
\usepackage{hyperref}
\bibpunct{(}{)}{;}{a}{}{,} 

\usepackage{siunitx}
\DeclareSIUnit{\Msun}{\mathrm{M}_\sun}
\DeclareSIUnit{\Zsun}{\mathrm{Z}_\sun}
\DeclareSIUnit{\comoving}{com.}
\DeclareSIUnit{\pc}{pc}
\DeclareSIUnit{\hubble}{\textit{h}}
\DeclareSIUnit{\codelength}{\comoving\kilo\pc\per\hubble}
\DeclareSIUnit{\yr}{yr}
\DeclareSIUnit{\dex}{dex}
\DeclareSIUnit{\erg}{erg}
\DeclareSIUnit{\sfrdunit}{\Msun\per\yr\per\cubic\mega\pc}
\title{Metal Enrichment by the First Stars Exploding at the Lower Energy Limit of Pair-Instability Supernovae}
\author{
Aron Kordt\inst{\ref{inst:ita}}\corrauth{aron.kordt@stud.uni-heidelberg.de} 
\and Simon C.\ O.\ Glover\inst{\ref{inst:ita}}\email{glover@uni-heidelberg.de}
\and Ralf S.\ Klessen\inst{\ref{inst:ita},\ref{inst:iwr}}\email{klessen@uni-heidelberg.de}
}

\institute{%
Universit\"{a}t Heidelberg, Zentrum f\"{u}r Astronomie, Institut f\"{u}r Theoretische Astrophysik, Albert-Ueberle-Str.\ 2, 69120 Heidelberg, Germany\label{inst:ita}
\and Universit\"{a}t Heidelberg, Interdisziplin\"{a}res Zentrum f\"{u}r Wissenschaftliches Rechnen, Im Neuenheimer Feld 225, 69120 Heidelberg, Germany\label{inst:iwr}
}

\begin{document}
\abstract{
	The first generation of stars, Population III (Pop III), is believed to be massive, with some potentially having masses in the range \qtyrange{140}{270}{\Msun} and capable of exploding as a pair-instability supernova (PISN). 
	Such events release large amounts of energy and produce substantial quantities of metals, suggesting that they should leave characteristic signatures in the abundance patterns of extremely metal-poor (EMP) stars observed in the local Universe. 
	No clear imprint of PISNe is seen in the local EMP star population, implying either that these events were rare or that stars forming from PISN-enriched gas had metallicities too high to find them in the EMP population. Previous work explored the latter possibility by investigating the enrichment by PISNe with masses and explosion energies at the upper end of the theoretical range (\qty{270}{\Msun}, \qty{e53}{\erg}).
	Here, we complement that work at the opposite extreme: Pop III stars at the lower mass (\qty{140}{\Msun}) and explosion energy (\qty{5e51}{\erg}) limit.
	Using a cosmological hydrodynamic simulation, we self-consistently track 
	the formation of Pop III stars, their radiative and mechanical feedback, and the subsequent formation of second-generation stars in metal-enriched gas. We find that all second-generation stars are exclusively internally enriched by their progenitor within the same halo, thereby imprinting the abundance pattern of a single first-generation star. The median [\element{Fe}/\element{H}]\ abundance of second-generation stars is $\sim -5.5$ which is \qty{2.9}{\dex} smaller than in the high-energy PISN case. 
	Our results demonstrate that if Pop III PISNe were common, we would expect to find stars with the characteristic odd-even abundance pattern produced by PISNe within the observed EMP population. Their absence in observations therefore strongly disfavours PISNe as the dominant channel of early metal enrichment.
}
\keywords{
	hydrodynamics -- 
	stars: Population III -- 
	stars: abundances -- 
	supernovae: general --
	galaxies: high-redshift -- 
	(cosmology:) early Universe
}

\maketitle
\nolinenumbers
\section{Introduction}

In the favoured standard cosmological model ($\Lambda$CDM), after the Big Bang the Universe consisted of hydrogen, helium, and small traces of deuterium and lithium \citep{GalliPalla2013}.
As a consequence, the first stellar generation, also called Population III (Pop III), formed from this pristine gas. 
Once the first stars explode as supernovae, their nucleosynthetic products are released into the interstellar and intergalactic medium (ISM and IGM, respectively).
Chemical evolution starts, and, once the metal-enriched gas collapses, stars form that are qualitatively more similar to stars in the present-day Universe. 

In the absence of metals, the first stars are distinct from the stars observed in the present-day Universe.
During star formation, cooling is dominated by \element[][][][2]{H} and \element{HD}, and the gas reaches a minimum temperature of $\sim\qty{200}{\kelvin}$, significantly higher than the \qtyrange{10}{20}{\kelvin} temperatures characteristic of present-day star-forming gas.
Many theoretical studies or simulations have been carried out to understand the fragmentation of primordial collapsing gas clouds to constrain the mass range and the initial mass function (IMF) of the first stars (see e.g.\ \citealp{Clark+2011}, \citealp{Greif+2011}, \citealp{StacyBromm2013}, \citealp{Hirano+2015}, \citealp{Jaura+2020}, or \citealp{KlessenGlover2023} for a recent review).
The specific evolution is a complex interplay of chemistry, radiation, magnetic fields and turbulence \citep{Clark+2011a,ShardaMenon2025,Sharda+2025,Mayer+2025}.
A consensus is that Pop III stars are on average more massive than present-day stars making two types of supernovae possible that mark the end of the stellar life \citep{KlessenGlover2023}.
Massive stars of typically \qtyrange{13}{40}{\Msun} are thought to undergo core collapse and release an energy of $\sim\qty{e51}{\erg}$ before a neutron star or black hole forms \citep{Costa+2025}.
Very massive stars of \qtyrange{140}{270}{\Msun} explode as pair-instability supernovae (PISN), producing large amounts of metals (up to \qty{60}{\Msun}), releasing up to \qty{e53}{\erg} of energy and leaving no remnant behind \citep{HegerWoosley2002,Takahashi+2018}.
Stars with masses between \qtyrange{40}{140}{\Msun} either collapse directly to black holes, or first explode as low-energy pulsational PISN, producing little or no metal enrichment \citep{Woosley2017}.

Direct detection of the first stars is extremely difficult, owing to the high redshift at which they form and the small size of the resulting stellar populations \citep[see e.g.][]{Zackrisson+2024}. 
Consequently, most of the constraints that we have on their properties come from indirect probes, primarily stellar archaeology \citep{Beers+2005,FrebelNorris2015,Bonifacio+2025}. 
In particular, extremely metal poor (EMP) stars with $[\element{Fe}/\element{H}] \le -3$ are widely used as probes of the properties of the earliest stellar generations, under the assumption that the elements in these stars were produced by one or a few SNe. PISN yields are expected to carry a characteristic odd-even abundance pattern \citep{HegerWoosley2002,Takahashi+2018}, and hence it should be easy to distinguish EMP stars enriched by PISNe from those enriched by core-collapse SNe. However, none of the EMP stars that have so far been found in the local Universe show unambiguous signs of enrichment by PISN; the few candidate objects that have been discovered have on further investigation typically been found to have abundance patterns more consistent with core-collapse SNe or SN~Ia than PISNe \citep[see e.g.\ the discussion in][]{Bonifacio+2025}. How we should interpret this apparent lack of stars enriched by PISN is unclear. Does it indicate that Pop III stars typically did not become massive enough to explode as PISN? Or are the stars formed from metal-rich PISN ejecta too metal rich to be found amongst the EMP stars, as a number of authors \citep[see e.g.][]{Karlsson+2008,Greif+2010,Whalen+2013,deBennassuti+2017,Salvadori+2019} have previously suggested?

To investigate the suggestion that by focussing on EMP stars, we might miss most stars enriched by Pop III PISN, \citet{Magg+2022} carried out a cosmological hydrodynamical simulation in which they explicitly tested the metal enrichment potential of single Pop III stars at the upper end of the PISN mass range (\qty{270}{\Msun}). These stars have the largest possible explosion energy (\qty{e53}{\erg}) and iron yield (\qty{56}{\Msun}) for this class of SNe. They found that although some second generation stars in their simulation formed with metallicities $[\element{Fe}/\element{H}] > -3$, and hence could be missed in any survey focussing on EMP stars, the majority of their second generation stars had metallicities placing them in the EMP regime. \citet{Magg+2022} therefore concluded that the rarity of low metallicity stars with signatures of PISN enrichment indicates that PISNe themselves were rare events in the early Universe. However, there remains the concern that this result may be a consequence of the decision by \citet{Magg+2022} to focus on the most energetic PISNe. Because of their high energies, these sweep up very large masses of gas, leading to substantial dilution of the metals they produce due to mixing within the supernova remnants. It is not immediately obvious that much lower energy PISNe that sweep up far less gas would produce a similar range of metallicities.

We complemented the study by \citet{Magg+2022} with a new simulation using PISNe with a mass (\qty{140}{\Msun}) and explosion energy (\qty{5e51}{\erg}) at the bottom of the PISN mass range. To allow for the simplest possible comparison with the results of \citet{Magg+2022}, we change only the supernova properties -- the PISN explosion energy and the metal yield -- and otherwise keep the initial conditions and simulation parameters identical.

Throughout this paper, we use these conventions.
Comoving units are indicated by \enquote{com}.
We use the cosmological parameters from \citet{PlanckCollaboration+2016} with $H_0=\qty{67.74}{\kilo\metre\per\mega\pc\per\second} = h \cdot \qty{100}{\kilo\metre\per\mega\pc\per\second}$, $\Omega_\Lambda=0.6911$, $\Omega_\textrm{b}=0.04864$, $\Omega_\textrm{m}=0.3089$, $\sigma_8=0.8159$, $n_\textrm{s}=0.961$.
When we refer to low energy PISNe, we mean PISNe at the lower explosion energy limit (\qty{5e51}{\erg}) while high energy PISNe refer to PISNe at the upper limit ($\qty{e53}{\erg}$; \citealt{HegerWoosley2002}).

Our paper is structured as follows.
Since this work is complementary to \citet{Magg+2022}, we briefly summarise the adopted models and methods in Section~\ref{sec:methods} but otherwise refer to the original paper for details.
In Section~\ref{sec:results}, we describe the results, starting from large scales and zooming into small scales while focusing on the time interval from the supernova explosion until a second-generation star forms. The results are set into context in Section~\ref{sec:discussion}.
There, we also discuss the implications of our results for Pop III stars.
Our main findings and conclusions are summarised in Section~\ref{sec:conclusion}.

\section{Methods} \label{sec:methods}
In our simulation, we self-consistently model the environments in which the first stars form, including the cosmological context, the collapse of gas clouds, and the feedback of the stars, both radiative and mechanical. We evolve the dynamics until the star-forming halos recollapse and a second-generation star forms. Our simulation setup is the same as in \citet{Magg+2022} except for the explosion energy and metal yield of the supernovae. We briefly summarize the key aspects here and refer the reader to \citet{Magg+2022} for a more detailed discussion.

\subsection{Simulation Code and Physical Model}
As the only difference in our setup compared to \citet{Magg+2022} concerns the properties of the SNe, the evolution prior to the explosion of the first Pop III SN should be identical. Therefore, rather than starting from the same initial conditions as in \citet{Magg+2022}, we took as our initial conditions a snapshot from their simulation at a redshift of $z \sim 23.79$, shortly before the formation of the first Pop III star. For completeness, we note here that the initial conditions of the original \citet{Magg+2022} simulation were 
generated with MUSIC \citep{HahnAbel2011} using the cosmological parameters from \citet{PlanckCollaboration+2016}.
The simulation volume is a cube of \qty{1}{\comoving\cubic\mega\pc\per\cubic\hubble} containing $512^3$ dark matter particles and initially the same number of gas cells. 
Therefore, the mass of a dark matter particle is $\sim \qty{790}{\Msun}$ and the initial mass of the gas cells is $\sim \qty{148}{\Msun}$. 
The gas cells received an initial kick velocity of \qty{4.9}{\kilo\meter\per\second}, following \citet{Schauer+2019}, which corresponds to the most likely value of the baryonic streaming velocity at redshift $z=200$ \citep{TseliakhovichHirata2010}, the redshift at which the initial conditions were initialised. For simplicity, we oriented the coordinate system so that this kick was in the positive $x$ direction.

The dynamic evolution is computed with the moving mesh code AREPO \citep{Springel2010,Pakmor+2016,Weinberger+2020}.
The code tessellates the domain with a space-filling Voronoi mesh. 
As such, it combines the Eulerian and Lagrangian formulation of hydrodynamics by calculating mass fluxes across cell boundaries and allowing the mesh-generating points to move. By default, AREPO refines or de-refines gas cells as necessary to keep their masses within a factor of two of their initial mass, but it also allows one to impose additional refinement criteria. In our simulations, we ensure that the code satisfies the \citet{Truelove+1997} criterion and resolves the Jeans mass by requiring that the local Jeans length be resolved by at least 8 cells. We also follow \citet{Magg+2022} and require that the Strömgren sphere \citep{Stromgren1939} surrounding each of our massive stars is resolved with at least 1000 cells. In addition, we limit the maximum effective cell diameter\footnote{Defined as $D = 2 (3 V / 4\pi)^{1/3}$, where $V$ is the volume of the cell.} to 0.15 pc for all cells within 10 pc of a star to ensure that we resolve the Sedov-Taylor radius of the PISN that they produce. Finally, to prevent cells close to the stars from becoming very small, we prevent their comoving volume from decreasing below $\qty{0.1}{\comoving\cubic\pc\per\cubic\hubble}$.

We include primordial non-equilibrium chemistry of \element{H}, \element{D}, \element{He}, \element[][][][2]{H}, \element{HD}, and their ions \citep{Clark+2011a}. A chemical network with primordial species is sufficient because we do not reach metallicities and densities such that metal cooling becomes important \citep{Jappsen+2007}.
The abundance of the elements \element{H}, \element{He}, and \element{D} are taken to be constant throughout the simulation.
Radiative transfer is implemented using SPRAI \citep{Jaura+2018,Jaura+2020}. 

\begin{table}
	\centering
	\caption{Energy bands used in our radiative transfer model. We also list the band-averaged cross-sections for the listed processes and the photon flux in each band produced by a single massive star.    \label{tab:photonscrosssec}}
	\begin{tabular}{ll@{}c @{\hspace{1em}}c}
		\toprule
		Energy [\unit{\electronvolt}] & Process & {$\sigma\,[\qty{e-18}{\square\centi\metre}]$} & {$\dot{N}\,[\qty{e49}{\per\second}]$} \\
		\midrule
		11.2 -- 13.6 & \element[][][][2]{H} dissociation    & 2.47 & 2.87\\
		13.6 -- 15.2 & \element[][][][2]{H} dissociation    & 2.47 & 1.93\\
		& \element{H} ionization               & 5.34 & \\
		15.2 -- 24.6 & \element[][][][2]{H} ionization    & 6.48 & 10.2\\
		& \element{H} ionization               & 2.23 & \\
		24.6 -- 136  & \element[][][][2]{H} ionization      & 1.57 & 14.9\\
		& \element{H} ionization               & 0.43 & \\
		& \element{He} ionization              & 4.13 & \\
		\bottomrule
	\end{tabular}
\end{table}

Stars are represented by sink particles \citep{Tress+2020}. If a gas cell exceeds a number density of \qty{e4}{\per\cubic\centi\metre}, we first check whether it is located at a local minimum of the gravitational potential, and then whether the gas within the sink formation radius $r_{\rm sink}$ is converging, collapsing and gravitationally bound. We also verify that there is no other sink particle within a distance $r_{\rm sink}$ of the cell. Cells that pass all of these checks are converted into collisionless sink particles. For the sink formation radius, we adopt a value $r_{\rm sink} = 2$~pc, corresponding approximately to the Jeans length at a density \qty{e4}{\per\cubic\centi\metre} and temperature $T=\qty{200}{\kelvin}$ that roughly correspond to the minimum turning point in the effective equation of state \citep{Yoshida+2006}. Sink particles can accrete mass if the density exceeds the sink formation density within the sink formation radius, provided that the gas is gravitationally bound to the sink. 

For the purposes of the numerical experiment presented in this paper, we assume that every sink produces the same amount of radiative feedback, regardless of its mass. 
We treat the sinks as \qty{200}{\Msun} zero-age main-sequence stars, using the luminosity from \citet{Schaerer2002} and assuming a black body spectrum in order to determine the photon flux in each of our radiative energy bins. The photon flux in each bin\footnote{Note that the values for the photon fluxes listed here are different from the ones listed in Table 1 in \citet{Magg+2022}. This is because of a typographical error in that paper: the values they quote in their table are all a factor of $\pi$ smaller than the values actually used in the code. Our Table~\ref{tab:photonscrosssec} gives the corrected values.}  and the bin-averaged cross-section for each chemical process (H$_{2}$ dissociation, H ionisation, H$_2$ ionisation, He ionisation) are listed in Table~\ref{tab:photonscrosssec}. We adopt a fixed stellar lifetime of \qty{2.2}{\mega\yr} and assume that the luminosity remains constant throughout this lifetime. Note that the mass of the star that we adopt for the purposes of computing its radiative feedback is not consistent with the mass that we assume for computing its mechanical feedback and metal enrichment (140~M$_\sun$). 
This inconsistency is due to the fact that we have deliberately adopted the same radiative feedback parameters (ionising photon fluxes and stellar lifetime) as in \citet{Magg+2022}, in order to ensure that any difference in outcome compared to that simulation can be ascribed to the difference in the supernova energy input and not differences in the details of the radiative feedback implementation.

At the end of the adopted stellar lifetime, each sink particle produces a single supernova explosion. 
We represent the energy input from this supernova by injecting thermal energy into the simulation with a total value of $E_\mathrm{SN}=\qty{5e51}{\erg}$. 
This explosion energy is the lowest amount that we expect a metal-free pair-instability supernova to produce \citep{HegerWoosley2002,Chen+2014,Takahashi+2018}, corresponding to the outcome for a 140~M$_\sun$ progenitor. 
For comparison, \citet{Magg+2022} instead inject an energy of $\qty{e53}{\erg}$, the highest energy that we expect for a metal-free pair-instability supernova, corresponding to a \qty{260}{\Msun} progenitor.
We insert this energy into the 1000 cells closest to the sink particle. 
Thanks to the high spatial resolution close to the sink owing to our Str\"omgren sphere refinement criterion (see Fig.~\ref{fig:resolution}), these cells occupy a volume much smaller than that of the supernova remnant at the end of its Sedov-Taylor phase, allowing us to resolve this phase in its evolution and preventing overcooling. 
At the same time as we inject the energy from the supernova, we also insert \num{e5} Monte Carlo tracer particles \citep{Genel+2013} into the same volume. 
These particles trace the flow of gas affected by the supernova feedback and allow us to identify the gas enriched by metals produced by the supernova.

The snapshot interval of $\sim \qty{0.2}{\mega\yr}$ is chosen to capture the lifetime of a star with roughly ten snapshots.
The simulation covers a total of \qty{55}{\mega\yr}, from an initial redshift $z\approx 23.79$ to a final redshift $z\approx 18.87$. However, towards the end of the simulation, we found that the radiative transfer portion of the calculation was dominating the runtime to an extreme extent, making it challenging to continue the simulation. To avoid this problem, we switched off radiative feedback from Pop III stars at a redshift $z\approx 19.57$, approximately \qty{10}{\mega\yr} before the end of the simulation. This allowed us to continue to self-consistently follow the evolution of metals injected by supernovae exploding at redshifts higher than this, which would otherwise have been computationally impractical, but has the consequence that any supernovae exploding after this time do so in an environment in which the effects of stellar feedback are not properly followed. For this reason, in this study we focus primarily on the fate of the metals produced by PISN exploding at $z > 19.57$. This computational choice is discussed further in Section~\ref{sec:discussion:shortcomings} below.
Our final decision to stop the simulation at $z\approx 18.87$ was motivated by its increasing memory demands, due to the heavy grid refinement applied around each sink particle. The total computational cost of the simulation was 
approximately 3.1 million core hours.

\subsection{Postprocessing}
The simulation was post-processed with the phase-space halo finder ROCKSTAR \citep{Behroozi+2013}.
This code uses the friends-of-friends algorithm to identify clusters of dark matter particles by linking particles to groups that are within a distance $b=0.24$ times the mean particle separation.
Although we can resolve a star-forming dark matter minihalo of a virial mass of \qty{e5}{\Msun} with at least 125 dark matter particles, the halo location and virial radius of such small halos are strongly affected by statistical fluctuations. In practice, however, the minihalos hosting Pop III star formation in our simulation were considerably more massive and were typically represented by $\sim 1000$ or more dark matter particles (see Table~\ref{tab:summary}), implying that their properties are well resolved 
\citep{Sasaki+2014}. We manually and iteratively optimized the ROCKSTAR parameters to yield a sufficiently smooth mass history for the main progenitor branch of the star-forming halos.
The halo catalogue was passed to CONSISTENT TREES \citep{Behroozi+2013a} to build dark matter merger tree.

The metal enrichment is traced with Monte Carlo tracer particles.
The number of tracer particles within a specified volume is converted to [Fe/H] with
\begin{equation}
	\label{eq:feh_abundance}
	[\element{Fe}/\element{H}] = \log \left(\frac{\textrm{\#~tracer} \cdot m_\textrm{Fe,SN}}{10^5 \cdot m_\textrm{gas} \cdot X} \frac{m_{\element{H}}}{m_{\element{Fe}}} \right) - \varepsilon_{\textrm{Fe},\sun}.
\end{equation}
Here, $m_\textrm{gas}$ and \#~tracer refer to the gas mass and number of tracer particles within the volume, respectively, 
$m_{\element{H}}$ and $m_{\element{Fe}}$ are the masses of a \element{H} and \element{Fe} atom, respectively,
$X=0.76$ is the primordial hydrogen mass fraction, and $\varepsilon_{\textrm{Fe},\sun}=-4.5$ is the logarithmic solar iron abundance with respect to hydrogen \citep{Asplund+2009}. 
For our assumed progenitor mass of \qty{140}{\Msun}, the value we adopt for the iron yield per supernova is $m_\textrm{Fe,SN}\approx\qty{5e-2}{\Msun}$ \citep{Nomoto+2013}, although we note that there is an uncertainty in the exact yields that one should expect from low energy PISN; see Section~\ref{sec:discussion:shortcomings} for a discussion of this point. Since all our PISNe have identical mass and energy, our metallicity results are easily scaled to other yield tables or other quantities using Eq.~\eqref{eq:feh_abundance}, e.g.\ the total metallicity is
\begin{equation}
	\label{eq:total_metallicity}
	\log Z = [\element{Fe}/\element{H}] + \varepsilon_{\textrm{Fe},\sun} - \log \frac{f m_{\element{H}}}{X m_{\element{Fe}}}\;,
\end{equation}
where $f=m_\textrm{Fe,SN}/\left(\sum_{A>\element{He}} m_{A,\textrm{SN}}\right)$ is the iron fraction relative to all metals using $m_{A,\textrm{SN}}$ as the supernova yield mass of element $A$.
In the case of our adopted model, $\log Z \approx [\element{Fe}/\element{H}] - 0.61$.

\section{Results} \label{sec:results}

During our simulation, we observe star formation in the same set of 21 halos discussed by \citet{Magg+2022}. Selected properties of these halos are summarized in Table~\ref{tab:summary}; c.f.\ Tables 1 and 3 in \citet{Magg+2022}, and note that we use the same halo numbering as in that work. In 12 of these halos, a PISN explodes during the period in our simulation in which we are still self-consistently following the effects of radiative feedback. In the remaining 9 halos (nos.\ 12--20 in the table), the first SN occurs only after the radiative feedback has been switched off, and so we do not consider these halos in our further analysis. 

Out of the 12 halos in which radiative feedback is followed self-consistently, a subset of 8 form a second generation star before the end of our simulation. In these halos, the fuel for this star formation is provided by the recollapse of gas enriched by the initial supernova, and for this reason we term this set of halos the \textit{recollapse} halos. 
The halos that host a supernova which explodes towards the end of the simulation do not recollapse, and so it could be that if we were to continue our simulation for longer, gas would also recollapse in some or all of these halos; however, this is speculative. 
In the halo with id 11, there are two stars forming at different locations in the same halo just \qty{13.8}{\kilo\yr} apart. 
Each of them is treated as a first star and only the next star forming in metal-enriched gas in this halo is treated as a second generation star. 
The simulation volume is too small to serve as a statistically representative, cosmological volume. Thus the exact number of star-forming halos is not a statistically robust quantity. Nevertheless,
the cosmological initial conditions provide a realistic environment for studying enrichment from Pop. III stars.

\begin{table*}
	\centering
	\caption{
		Properties of the star-forming halos.
		$z_{\rm col}$ indicates the redshift at which the first star forms and $z_{\rm sn}$ the redshift at which the subsequent PISN explodes. $z_{\rm rec}$ indicates, when appropriate, the formation redshift of the second generation star. We also list the virial mass of the halo at $z_{\rm col}$ and $z_{\rm rec}$, the recollapse time $t_{\rm rec}$ (i.e.\ the time delay between $z_{\rm sn}$ and $z_{\rm rec}$), the mass of dense gas in the halo at $z_{\rm sn}$, the metallicity of the second generation star that forms and the average metallicity of the halo at the same time. Finally, the column denoted ``flag'' classifies the halos into good (radiative transfer until the supernova of the first-generation star), no rec.\ (halo does not recollapse) and no RT (radiative transfer was switched off during or before the first star forms).
	}
	\label{tab:summary}\begin{tabular}{l@{}S@{\hspace{1em}}S@{\hspace{1em}}S@{\hspace{1em}}S@{\hspace{1em}}S@{\hspace{1em}}S@{\hspace{1em}}S@{\hspace{1em}}S@{\hspace{1em}}S@{\hspace{1em}}S@{\hspace{1em}}S@{\hspace{1em}}}
		\toprule
		{halo} & {$z_\textrm{col}$} & {$z_\textrm{sn}$} & {$z_\textrm{rec}$} & {$M_\textrm{vir,col}\,[\qty{e6}{\Msun}]$} & {$M_\textrm{vir,rec}\,[\qty{e6}{\Msun}]$} & {$t_\textrm{rec}\,[\unit{\mega\yr}]$} & {$M_\textrm{dens}\,[\qty{e3}{\Msun}]$} & {$[\element{Fe}/\element{H}]_\textrm{2nd}$} & {$[\element{Fe}/\element{H}]_\textrm{halo}$} & {flag} \\
		\midrule
		0 & 23.90 & 23.64 & 22.03 & 1.50 & 2.42 & 15.09 & 1.36 & -6.21 & -5.41 & {good} \\
		1 & 22.26 & 22.04 & 20.15 & 0.62 & 1.05 & 21.35 & 7.36 & -6.17 & -5.26 & {good} \\
		2 & 21.82 & 21.61 & 21.59 & 0.83 & 0.87 & 0.23 & 10.83 & -5.11 & -4.73 & {good} \\
		3 & 21.79 & 21.59 & 20.57 & 0.83 & 1.06 & 11.44 & 13.50 & -5.06 & -4.92 & {good} \\
		4 & 20.45 & 20.27 & 18.95 & 0.81 & 1.39 & 17.79 & 15.99 & -5.97 & -5.06 & {good} \\
		5 & 20.44 & 20.27 & {---} & 0.43 & {---} & {---} & 0.03 & {---} & {---} & {no rec.} \\
		6 & 20.00 & 19.83 & 19.55 & 0.88 & 1.02 & 3.80 & 12.76 & -4.81 & -4.83 & {good} \\
		7 & 19.96 & 19.80 & {---} & 0.93 & {---} & {---} & 0.23 & {---} & {---} & {no rec.} \\
		8 & 19.85 & 19.68 & {---} & 1.13 & {---} & {---} & 2.58 & {---} & {---} & {no rec.} \\
		9 & 19.79 & 19.63 & {---} & 0.79 & {---} & {---} & 5.01 & {---} & {---} & {no rec.} \\
		10 & 19.77 & 19.61 & 19.18 & 1.36 & 1.67 & 5.93 & 8.23 & -5.33 & -5.11 & {good} \\
		11 & 19.74 & 19.57 & 19.74 & 0.97 & 1.01 & 0.50 & 11.58 & -5.62 & -4.46 & {good} \\
		12 & 19.68 & 19.52 & {---} & 1.06 & {---} & {---} & 13.52 & {---} & {---} & {no RT} \\
		13 & 19.66 & 19.50 & {---} & 1.04 & {---} & {---} & 20.13 & {---} & {---} & {no RT} \\
		14 & 19.39 & 19.23 & {---} & 0.66 & {---} & {---} & 5.14 & {---} & {---} & {no RT} \\
		15 & 19.28 & 19.12 & {---} & 0.94 & {---} & {---} & 5.12 & {---} & {---} & {no RT} \\
		16 & 19.16 & 19.01 & {---} & 0.70 & {---} & {---} & 4.78 & {---} & {---} & {no RT} \\
		17 & 19.10 & 18.95 & {---} & 1.17 & {---} & {---} & 14.35 & {---} & {---} & {no RT} \\
		18 & 19.07 & 18.91 & {---} & 0.85 & {---} & {---} & 6.59 & {---} & {---} & {no RT} \\
		19 & 18.98 & 18.83 & {---} & 0.83 & {---} & {---} & {---} & {---} & {---} & {no RT} \\
		20 & 18.90 & 18.75 & {---} & 1.29 & {---} & {---} & {---} & {---} & {---} & {no RT} \\
		\bottomrule
	\end{tabular}
\end{table*}

\subsection{Resolution}
\begin{figure}
	\centering
	\includegraphics[width=\linewidth]{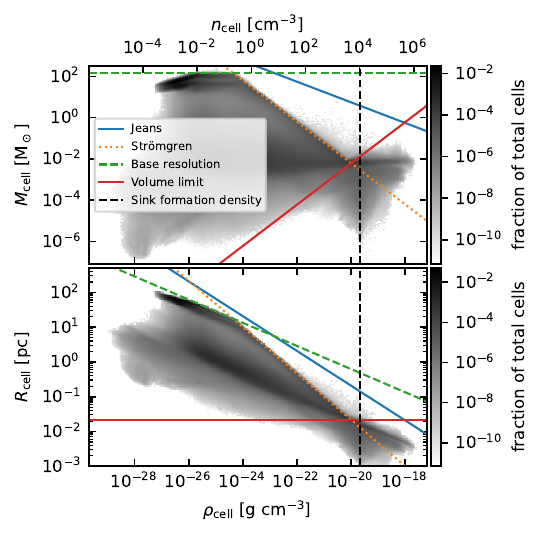}
	\caption{
		Time-averaged 2D histogram of the cell mass and cell radius as a function of the cell density.
		The Jeans criterion requires that the Jeans length is resolved by at least 8 cells, with the value of the Jeans length being computed assuming a temperature of \qty{200}{\kelvin}, approximately the lowest value that the gas reaches. For the 
		Strömgren criterion, we follow \citet{Magg+2022} and compute the size of an ideal Strömgren sphere at the density in question for a gas temperature of $10^{4} \: {\rm K}$ and an ionising source producing \qty{4e49} ionising photons per second. We then require that this Strömgren sphere is resolved with at least 1000 gas cells. In practice, the stars formed in our simulation have higher ionising photon luminosities, and so their Strömgren spheres will be even better resolved. We also indicate the minimum volume limit (horizontal red line) and the density above which sink formation is allowed (vertical dashed line).
		\label{fig:resolution}}
\end{figure}

The mass and spatial resolution of the gas cells averaged over time is shown in Fig.~\ref{fig:resolution}. At low densities, the cell mass is close to the base resolution of $\sim 125 \: {\rm M_{\odot}}$, indicated by the horizontal green dashed line. However, above $n \sim 1 \: {\rm cm^{-3}}$, the Str\"omgren refinement criterion dominates, reducing the cell mass with increasing density as $M_{\rm cell} \propto n^{-1}$ until it reaches a value of $\sim 0.01 \: {\rm M_{\odot}}$ at the sink creation density. Although the simulation also includes a Jeans length refinement criterion, this is less strict than the Str\"omgren refinement criterion because of its weaker scaling with density ($M_{\rm cell} \propto n^{-1/2}$).

It is also clear from Fig.~\ref{fig:resolution} that some cells end up with masses and/or volumes smaller than we would expect based on our refinement criteria and volume limit. This is a consequence of the 
hybrid Lagrangian-Eulerian scheme used in AREPO. The mesh-generating points of the cells move with the flow of the fluid to resolve high-density environments with a larger number of cells. However, if any of these cells move from higher densities back to lower densities, they can retain their small sizes temporarily, since de-refinement is not an immediate process, meaning that their masses can become very small. In addition, for reasons of numerical stability, AREPO limits the maximum change in volume that is allowed between neighbouring cells, so very low density cells close to high density cells can end up with much smaller volumes than one might initially expect.

\subsection{Large-scale Evolution}
\begin{figure}
	\centering
	\includegraphics[width=\linewidth]{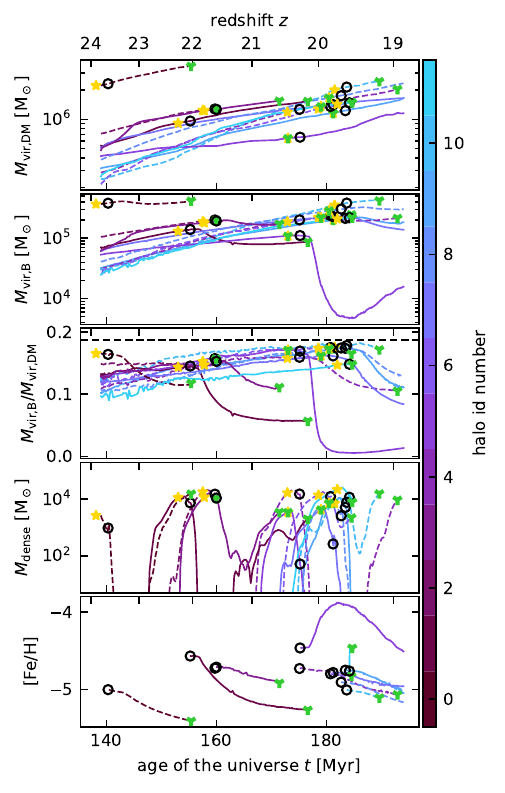}
	\caption{
		Evolution of the dark matter mass, the baryon mass contained within the virial radius, the baryon-to-dark matter ratio, the mass of dense gas having $n>\qty{e3}{\per\cubic\centi\metre}$ and the halo-averaged [\element{Fe}/\element{H}]\ in the star-forming halos as a function of time. 
		For each halo, we mark the time that the first star forms (yellow star), the time that the first supernova explodes (black circle) and, if appropriate, the time that the gas recollapses and forms the first metal-enriched star (green triangle).
		Lines are alternatingly solid and dashed for clarity.
		They are truncated at the time of recollapse or at the end of the simulation. 
		The horizontal black dashed line in the middle panel shows the cosmic average baryon-to-dark matter ratio.
	}
	\label{fig:halomass}
\end{figure}

The large scale evolution of the halos is driven by the gravitational interaction dominated by dark matter. 
For each star-forming halo, we compute the time evolution of the dark matter halo mass, the baryonic mass (i.e.\ the mass of gas contained within the virial radius of each halo), the baryon-to-dark matter mass ratio, the mass of dense gas (defined here as gas with number density $n>\qty{e3}{\per\cubic\centi\metre}$) and the halo average [\element{Fe}/\element{H}]\ abundance.
The results for the 12 halos that form stars before we switch off radiative transfer are shown in the top three panels in Fig. \ref{fig:halomass}. Each quantity is plotted until either a second generation star forms in the halo or until the end of the simulation in the case of the 4 halos that do not form second generation stars. 

The minimum dark matter halo mass at the onset of star formation is \qty{4.3e+05}{\Msun} (halo number 5), but most of the star-forming halos are more massive, with masses of \qtyrange[range-phrase={ -- }]{8e+05}{1.5e+06}{\Msun}; for comparison, the value predicted by the \citet{Schauer+2021} fitting function for our adopted streaming velocity is \qty{1.1e+06}{\Msun}.
At the time that the halos form their first stars, their baryon-to-dark matter ratio is appreciably smaller than the cosmic mean value, owing to the influence of the streaming velocity.
Since cosmic streaming is a second-order effect, its impact reduces at later cosmic times once gravitational collapse produces sufficiently massive objects to gravitationally bind the gas \citep{TseliakhovichHirata2010}.
We find a median baryon to dark matter fraction within the virial radius of the recollapsing halos at 0.155 with a minimum of 0.144 and a maximum of 0.175 compared to the cosmic mean $\Omega_\textrm{b}/\Omega_\textrm{dm}=0.188$.

Once the first Pop III stars explode as supernovae, they inject so much thermal energy into the gas surrounding the star that the blast wave expands well beyond the virial radius. This produces significant outflows in some halos, dramatically reducing their baryonic mass and baryon-to-dark matter ratio. In the most extreme case (halo number 5), the baryon-to-dark matter ratio is reduced from 0.171 at star formation to only 0.006, but this halo does not recollapse during the simulation. Among the recollapsing halos, the most gas-poor halo (halo number 1) has a baryon-to-dark matter ratio of 0.144 at the time of star formation, and yet, the halo forms a second generation star roughly \qty{21}{\mega\yr} after the SN explosion. In other cases, however, there is much less loss of gas and/or a much shorter delay between the SN and the formation of a second generation star. 

We observe a total of \qtyrange{e3}{e4}{\Msun} of dense gas, which is defined here as gas denser than $10^{3} \: {\rm cm^{-3}}$, at the time that the first star forms.
There is no significant difference in the total mass of dense gas when we apply additional restrictions on the gas temperature $<\qty{300}{\kelvin}$ or additionally the molecular hydrogen abundance $x_{\element[][][][2]{H}} > \num{e-4}$.
Once a Pop III star forms, the total mass of dense gas in the halo decreases because the UV radiation from the star heats the gas.
In eight halos, the supernova completely evaporates any dense gas structures.
In the remaining ones, dense gas structures remain, yet their total mass is reduced by \qty{2}{\dex}.

The bottom panel in Fig.~\ref{fig:halomass} shows the time evolution of the halo-averaged [Fe/H]. This initially has a value in the range \numrange{-5}{-4.5}, depending on the baryonic mass of the halo, but decreases with time as metal-enriched gas flows out of the halo, reaching a value at recollapse that is typically \qtyrange{0.5}{1}{\dex} smaller than the initial value.

\begin{figure*}
	\centering
	\includegraphics[width=18cm]{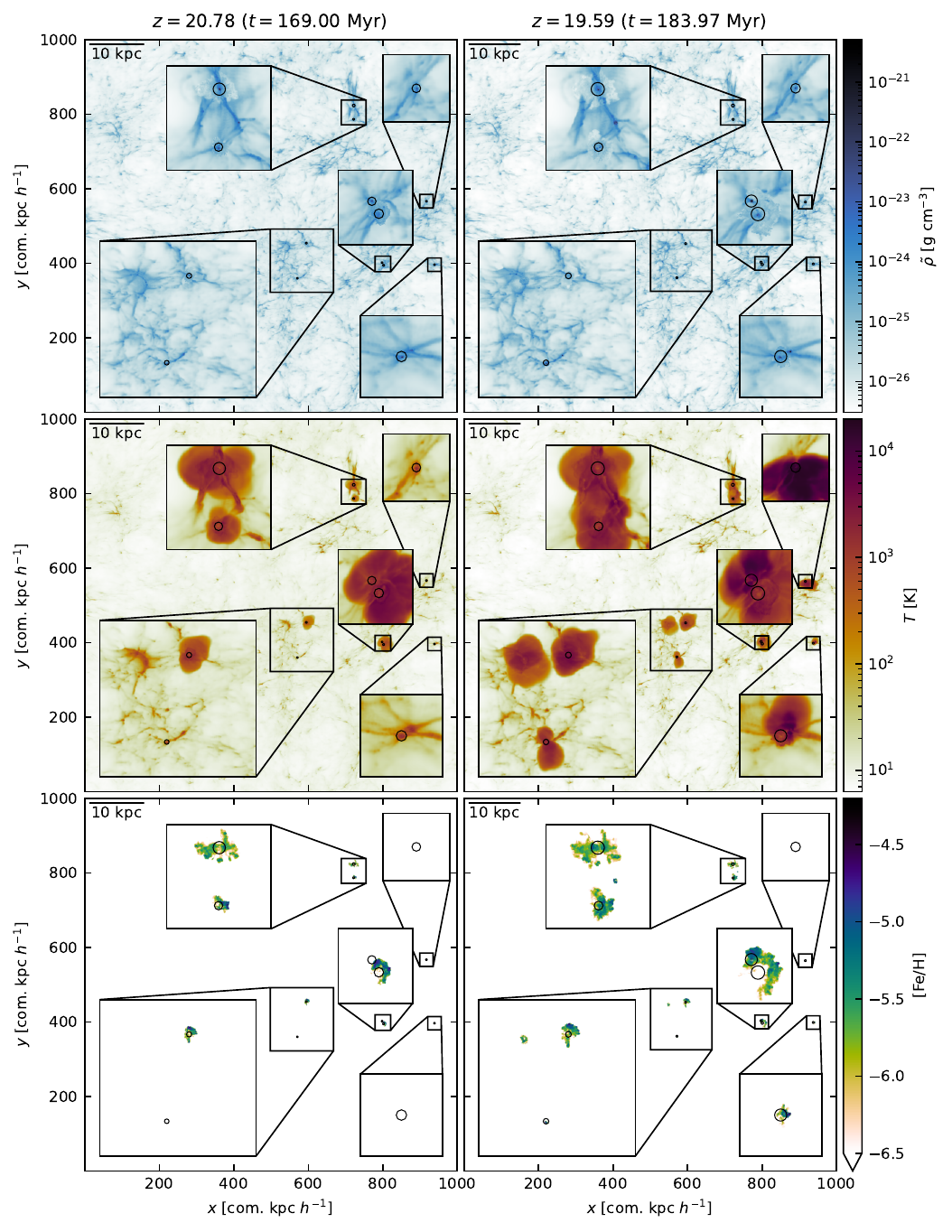}
	\caption{
		Projection of the distribution of gas density $\tilde{\rho}$ (see Eq.~\eqref{eq:densitysquared}; upper row), mass-weighted temperature (middle row) and [Fe/H] (see Eq.~\eqref{eq:feh_abundance}; bottom row).
		Circles indicate the virial radii of the star-forming halos. 
		Zoom insets show more details of the regions where these halos are located.
	}
	\label{fig:fullboxprojection}
\end{figure*}

In Fig.~\ref{fig:fullboxprojection}, projections of the full simulation box are shown at two times: at a redshift $z = 20.78$, 
approximately half-way through the simulation (left column) and at a redshift $z = 19.59$ (right column), which corresponds to the last snapshot that was output while radiative feedback was still switched on. The panels show the density-weighted gas density 
\begin{equation}
	\label{eq:densitysquared}
	\tilde{\rho}=\int \rho^2 \mathrm{d}z / \int \rho \mathrm{d}z,
\end{equation}
the mass-weighted temperature $\tilde{T}=\int T m \mathrm{d}z / \int m \mathrm{d}z$ and the mass-weighted [\element{Fe}/\element{H}]\ abundance projected along the z-axis.
Zoom-insets show details of the (future) star forming regions or the regions where stars already exploded in supernova explosions.
Additionally, the virial radii of star-forming halos are shown as circles.

In the projection of the gas density, the filamentary structure of the cosmic web is clearly visible.
Due to the projection, the information of depth is lost such that star-forming halos seem to cluster in specific regions. 
The cosmic streaming creates a smeared-out effect along the x-axis which is the axis of the relative velocity.

\subsection{Supernova Explosion and Enrichment}

The metal enrichment process driven by the supernova and their blast waves is non-isotropic.
Dense gas near the explosion site remains largely unaffected, while the blast wave propagates more rapidly through less dense regions.
Filamentary structures, which efficiently channel cold gas into the halo, are minimally impacted by the supernova shock, as their high-density environment slows the expansion of the blast wave.
While the metal enrichment from a single supernova can extend beyond the virial radius of its host halo and mix turbulently with the inter-halo medium, we emphasize that this process is not uniform in space (see zoom insets in Fig. \ref{fig:fullboxprojection}). 
Instead, the degree of metal enrichment varies due to differences in gas density and structural features. Nevertheless, it is still informative to look at how the characteristic metallicity of the region enriched by a single PISN evolves over time.

To do this, we compute for each PISN the maximum enriched radius as a function of the time since the supernova explosion ($\Delta t$). The maximum enriched radius is defined as the farthest distance from the explosion site where metal ejecta from a single supernova can be found. We also define the enriched mass to be the total gas mass enclosed in a sphere of this radius. This serves as an upper limit on the actual mass enriched by the supernova, and hence the mean metallicity of this enriched mass is a lower limit on the true mean metallicity.
We use this approach because the total number of tracer particles inserted in the supernova explosions turned out in practice to be too small to fully sample the metal-enriched ejecta: not every cell in metal-enriched regions hosts tracer particles, and the exact location of an individual tracer is not deterministic due to the stochastic nature of Monte Carlo tracer particles \citep{Genel+2013}. Therefore, determining the enriched mass by summing up the mass of only those cells containing tracer particles yields a lower limit on the enriched mass. 

\begin{figure}
	\centering
	\includegraphics[width=\linewidth]{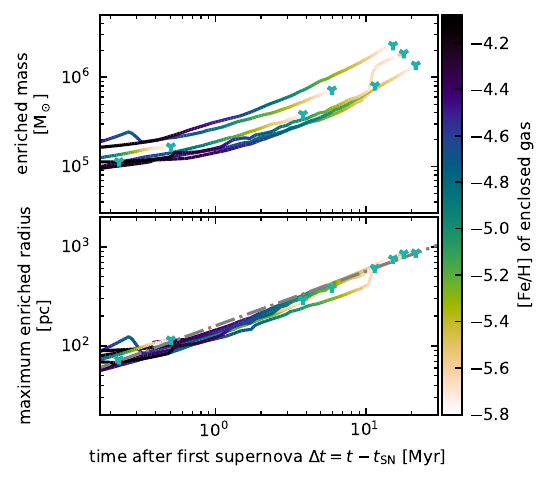}
	\caption{The upper panel shows the total mass that the supernova yields dilute into since the supernova explosion. The lower panel shows the maximum distance that the supernova yields reach since the supernova explosion. The lines are truncated at the time when a new star forms within the halo and the lines are colour-coded with the [Fe/H] abundance of the enriched mass or volume assuming spherical symmetry.}
	\label{fig:enrichment}
\end{figure}

In the upper panel of Fig. \ref{fig:enrichment}, we show the enriched mass as a function of time since the supernova explosion, while the lower panel illustrates the maximum enriched radius ($R_{\rm max}$) over the same time interval. We show this for each of the recollapse halos up to the point at which a second generation star forms (indicated by the turquoise symbol). 
In the following, we refer to the time delay between the SN explosion and the formation of the first second generation star as the \textit{recollapse time}, $t_\text{rec}$. In the Appendix, we show a version of this figure in which the enriched mass and maximum enriched radius are normalised by the virial mass and virial radius, respectively (Fig.~\ref{fig:enrichmentnorm}).

The enriched radius shows only small variations between the star-forming halos, but the enriched mass has somewhat larger variations spanning a range of about $\sim \qty{0.2}{\dex}$. Since all of the supernovae inject the same mass of metals, this difference is primarily driven by differences in the total mass of host halo and the distribution of mass within it. The average metallicity within $R_{\rm max}$ peaks at $[\element{Fe}/\element{H}]\sim\num{-4.1}$ just after the supernova but decreases quickly as the remnant expands and sweeps up additional mass, diluting the yields. In the two halos where recollapse happens rapidly after the SN ($t_{\rm rec} < 1 \: {\rm Myr}$), the mean metallicity at recollapse is $[\element{Fe}/\element{H}]\sim\num{-4.3}$ and $\sim\num{-4.5}$, respectively. In the other six halos that recollapse, the supernova blast wave has more time to penetrate the ambient gas, and the supernova yields dilute to $[\element{Fe}/\element{H}]\lesssim \num{-5.2}$.

We compare the expansion of the supernova blast wave with the Sedov-Taylor theory with power-law density variations at the time just before the supernova and at recollapse.
The density profile $\rho\propto r^{-k_\rho}$ outside the Strömgren sphere has the exponent $k_{\rho,\textrm{SN}}\approx\num{2}$ at the time just before the supernova explosion (see Fig.~\ref{fig:radialprofshells} in the Appendix), flattening slightly to $k_{\rho,\textrm{rec}}\approx\num{1.9}$ at later times.
In a power-law density profile, we would expect the shock front to propagate during the Sedov-Taylor phase according to $\log R_s\propto {2/(5-k_\rho)}\log t$ \citep{OstrikerMcKee1988} which is $\log R_s\propto \num{0.67} \cdot \log t$ at the time of the supernova explosion and $\log R_s\propto {\num{0.64}}\cdot \log t$ at the time of recollapse.
However, a power-law fit to the maximum enriched radius at the time of recollapse yields $\log \left(r/\unit{\pc}\right) = \num{0.57} \cdot \log \left(\Delta t / \unit{\mega\yr}\right)+ \num{2.20}$.

The Sedov-Taylor theory assumes that the blast wave expands into an isotropic medium with power-law density variations until it dissipates as a sound wave which appears in this case as a halo outflow. Since the ambient medium actually is non-isotropic and not at rest but gas is infalling into the halo, we observe that the blast waves propagate slower than theoretically expected. 
This demonstrates the need for accurate hydrodynamic modelling of metal enrichment by supernova explosions but this detailed hydrodynamic computation and its derived upper limits fit the simplified theory remarkably well.

Supernova explosions are among the most energetic events in the universe.
It is apparent from Fig.~\ref{fig:halomass} that a single PISN explosion is easily capable of removing a significant amount of gas from the inner region of the halos.
We compute the gravitational binding energy of the gas to the halo assuming it is at rest, thus effectively neglecting the additional work required to reverse any inward flow of the gas during the expansion.
The contribution of dark matter to the gravitational binding energy is evaluated assuming a Navarro-Frenk-White (NFW) profile \citep{Navarro+1997} using the fitted parameters from ROCKSTAR.
Since the NFW profile has a divergent mass, we only evaluate the binding energy up to the virial radius.
The typical binding energy is in the range $U\sim\qtyrange{5e50}{2.5e51}{\erg}$, which is \qtyrange{-1.0}{-0.3}{\dex} smaller than the explosion energy of the supernova. The single exception is halo number 0, whose binding energy is $\qty{9e51}{\erg}$ and hence is \qty{0.27}{\dex} larger than the supernova energy.
For comparison, the binding energy of the gas in these halos is roughly \qtyrange{-2.3}{-1.0}{\dex} smaller than the PISN explosion energy of \qty{e53}{\erg} adopted by \citet{Magg+2022}, implying that it is much easier for those high energy PISN to expel gas from the halo.

\subsection{Recollapse and Metallicity of Second Generation Stars}
\begin{figure*}
	\centering
	\includegraphics[width=18cm]{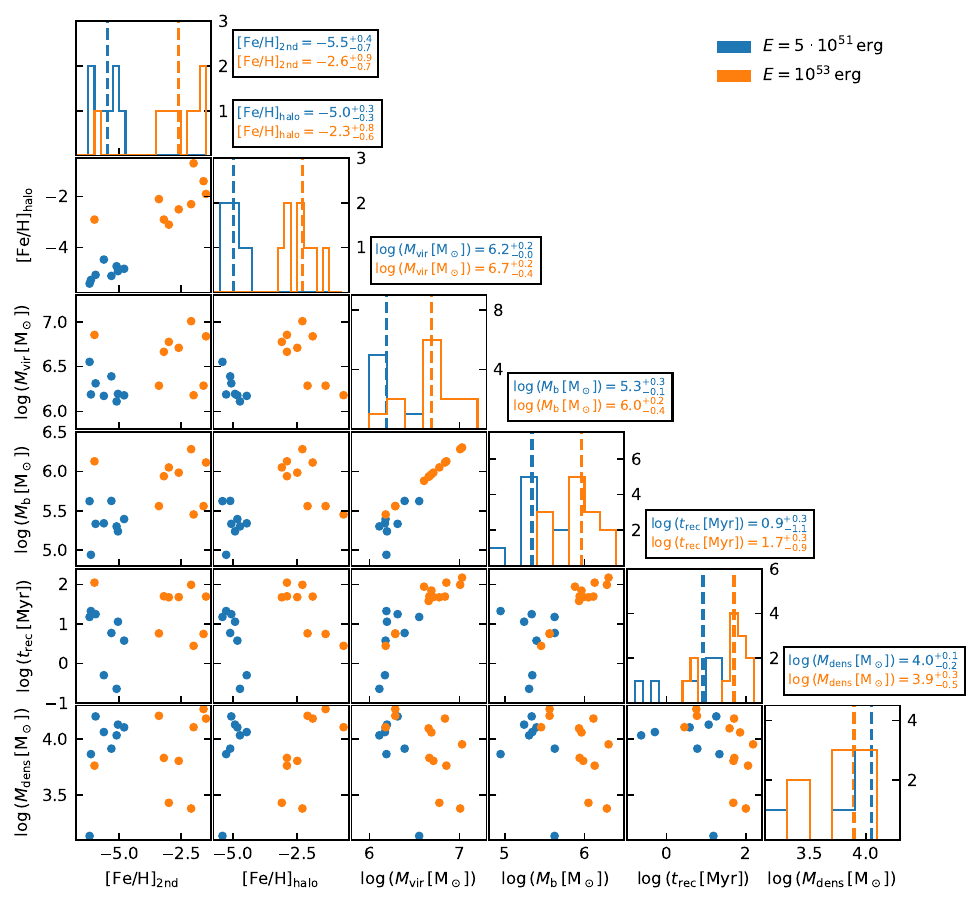}
	\caption{
		Pair-wise correlations in the lower-left triangle of the metallicity of the second generation stars
		that form ($[\element{Fe}/\element{H}]_\textrm{2nd}$), the mass-weighted mean
		metallicity of the halo ($[\element{Fe}/\element{H}]_\textrm{halo}$), 
		the dark matter virial mass $M_\textrm{vir}$, the baryonic mass within the virial radius $M_\textrm{b}$ (all computed at recollapse); the recollapse time $t_\textrm{rec}$; and the total mass of dense ($n>\qty{e3}{\per\cubic\centi\metre}$) gas $M_\textrm{dense}$ (computed at $z_{\rm sn}$).
		The panels on the diagonal show the respective distribution highlighting the median (50th percentile, vertical dashed lined). Values for the 16th, 50th and 84th percentiles are shown in the boxes next to the panels with the distributions. 
		The results from our study are shown in blue and the results from \citet{Magg+2022} for a much higher PISN explosion energy are shown in orange.
	}
	\label{fig:correlations}
\end{figure*}

Gravitational forces gradually reaccrete matter in the inner regions of the halos and stars form if the star formation criteria are met. In Fig. \ref{fig:correlations}, we show the relation between several different quantities at recollapse (see also Table \ref{tab:summary}): 
the metallicity of the second-generation stars ($[\element{Fe}/\element{H}]_\textrm{2nd}$) and the mass-weighted average value within the halo ($[\element{Fe}/\element{H}]_\textrm{halo}$); the virial mass $M_\text{vir}$; the baryonic mass content within the virial radius, $M_{\rm b}$; the recollapse time $t_\textrm{rec}$; and the mass of dense gas $M_\textrm{dens}$ having $n>\qty{e3}{\per\cubic\centi\metre}$ at the time of the supernova.
We also show the corresponding results from \citet{Magg+2022}. In Table~\ref{tab:correlations}, we list the Pearson correlation coefficients $r$ of these quantities, first distinguishing between our simulation and that by \citet{Magg+2022}, and then looking at the correlations in the joint dataset. 

\begin{table*}
	\centering
	\caption{
		Correlation coefficients.
		In the upper section, the upper right triangle uses the data from table 2 in \citet{Magg+2022} for PISNe exploding with $E=\qty{e53}{\erg}$, and the lower left triangle uses data from this work for PISNe exploding with $E=\qty{5e51}{\erg}$.
		In the lower section, both datasets are combined.
	}
	\label{tab:correlations}
	\begin{tabular}{ll@{}S@{\hspace{1em}}S@{\hspace{1em}}S@{\hspace{1em}}S@{\hspace{1em}}S@{\hspace{1em}}S@{\hspace{1em}}}
		\toprule
		& & {$[\text{Fe/H}]_\text{2nd}$} & {$[\text{Fe/H}]_\text{halo}$} & {$\log \left( M_\text{vir} \,[\text{M}_\odot] \right)$} & {$\log \left( M_\text{b} \,[\text{M}_\odot] \right)$} & {$\log \left( t_\text{rec}\, [\text{Myr}] \right)$} & {$\log \left( M_\text{dens}\, [\text{M}_\odot] \right)$} \\
		\midrule
		per group & {$[\text{Fe/H}]_\text{2nd}$} & {---} & 0.59 & -0.24 & -0.24 & -0.39 & 0.27 \\
		& {$[\text{Fe/H}]_\text{halo}$} & 0.59 & {---} & -0.70 & -0.70 & -0.80 & 0.66 \\
		& {$\log \left( M_\text{vir} \,[\text{M}_\odot] \right)$} & -0.53 & -0.72 & {---} & 1.00 & 0.93 & -0.36 \\
		& {$\log \left( M_\text{b} \,[\text{M}_\odot] \right)$} & 0.11 & -0.11 & 0.69 & {---} & 0.93 & -0.36 \\
		& {$\log \left( t_\text{rec}\, [\text{Myr}] \right)$} & -0.47 & -0.82 & 0.54 & -0.08 & {---} & -0.52 \\
		& {$\log \left( M_\text{dens}\, [\text{M}_\odot] \right)$} & 0.56 & 0.65 & -0.77 & -0.42 & -0.28 & {---} \\
		\midrule
		both & {$[\text{Fe/H}]_\text{2nd}$} & {---} & 0.88 & 0.37 & 0.53 & 0.23 & 0.13 \\
		& {$[\text{Fe/H}]_\text{halo}$} &  & {---} & 0.37 & 0.54 & 0.25 & 0.16 \\
		& {$\log \left( M_\text{vir} \,[\text{M}_\odot] \right)$} &  &  & {---} & 0.95 & 0.84 & -0.41 \\
		& {$\log \left( M_\text{b} \,[\text{M}_\odot] \right)$} &  &  &  & {---} & 0.72 & -0.33 \\
		& {$\log \left( t_\text{rec}\, [\text{Myr}] \right)$} &  &  &  &  & {---} & -0.40 \\
		& {$\log \left( M_\text{dens}\, [\text{M}_\odot] \right)$} &  &  &  &  &  & {---} \\
		\bottomrule
	\end{tabular}
\end{table*}

We see from Fig.~\ref{fig:correlations} that in the low energy PISN case examined in this paper, the metallicities of the second generation stars and their host halos are dramatically smaller than those in the \citet{Magg+2022} simulation, by around 2.5 orders of magnitude. In contrast to the high energy case, which produces some stars that are too metal enriched to be classified as EMP stars, the simulation with low energy PISNe produces only stars with $[{\rm Fe/H}] \ll 3$ that should therefore appear in surveys of EMP stars, provided that they exist in the real Universe.

In the low energy simulation, we find two groups of recollapse times:
a few halos that recollapse rapidly with $t_\textrm{rec}<\qty{1}{\mega\yr}$ and a larger group that recollapse more slowly with $t_\textrm{rec}\gtrsim\qty{4}{\mega\yr}$. This is similar to the pattern that \citet{Magg+2022} find, although the recollapse times in the low energy case are shorter by \qty{0.8}{\dex} on average. This result is easy to understand as a consequence of the difference in explosion energies. A lower energy PISN unbinds less gas from the halo and hence the enriched gas falls back into the halo more rapidly and takes much less time to reach the density required for star formation. This not only leads to a lower recollapse time, but also to the recollapsing halos having smaller $M_\textrm{vir}$ and $M_\textrm{b}$ at recollapse, since they have less time in which to grow.
In the high energy case, there is a strong correlation between $M_\textrm{vir}$ or $M_\textrm{b}$ and $t_\textrm{rec}$ having a Pearson correlation coefficient $r=0.93$.
In the low energy case, the correlation between $ M_\textrm{b}$ and $ t_\textrm{rec}$ breaks: the Pearson correlation coefficient $r$ is $r = -0.08$ 
while a moderate correlation with $r=0.54$ remains between $M_\textrm{vir}$ and $t_\textrm{rec}$.
In general, we cannot easily disentangle the influence of explosion energy on the recollapse time and halo masses, both dark matter and baryonic mass.

\begin{figure*}
	\centering
	\includegraphics[width=18cm]{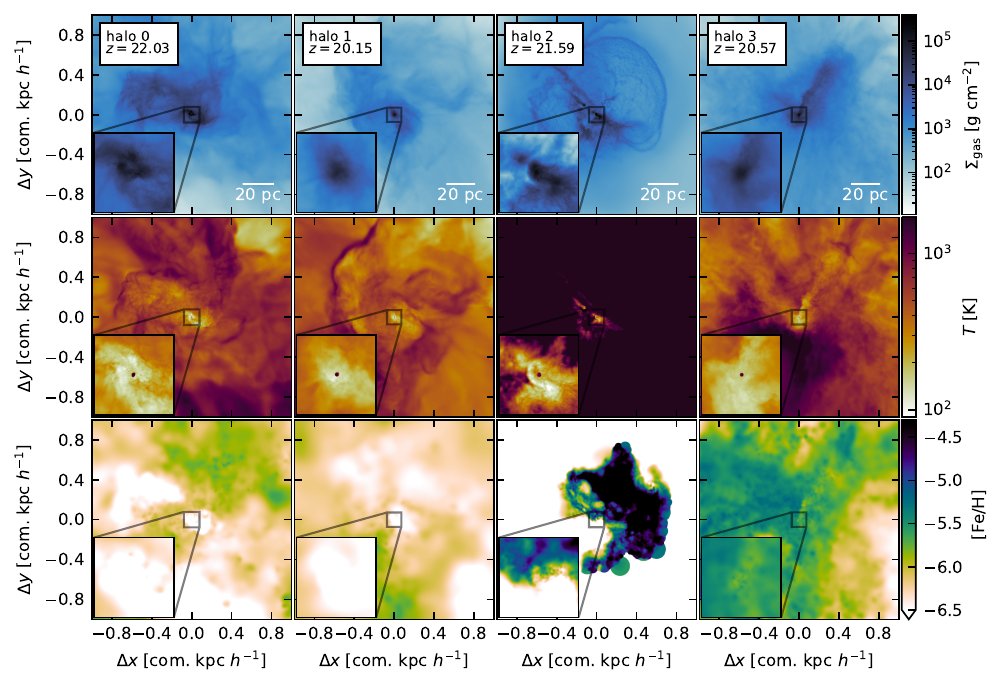}
	\caption{Projected density, mass-weighted temperature and mass-weighted [\element{Fe}/\element{H}] (top to bottom) for halos 0, 1, 2 and 3 (left to right) at the time of recollapse, i.e.\ just before a second-generation star forms. The projections are centred on the location at which that star will form.}
	\label{fig:sfbox0}
\end{figure*}

\begin{figure*}
	\centering
	\includegraphics[width=18cm]{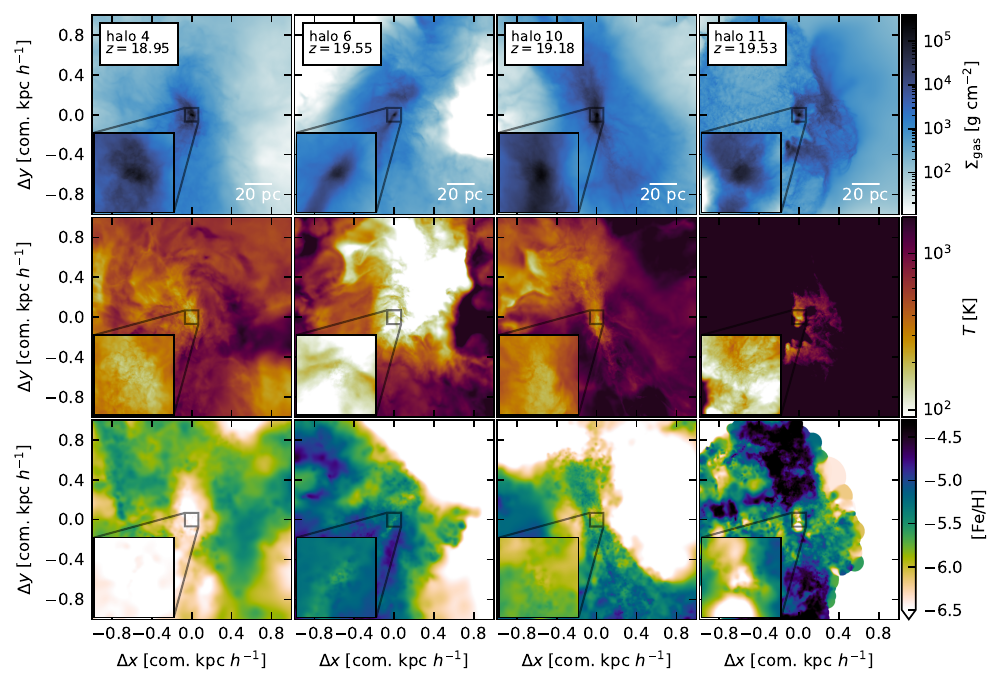}
	\caption{Same as Fig.~\ref{fig:sfbox0} but for halos 4, 6, 10 and 11.}
	\label{fig:sfbox1}
\end{figure*}

Looking more closely at the two halos in our simulation that have recollapse times $t_\textrm{rec}<\qty{1}{\mega\yr}$, we see that this is a consequence of dense gas structures in these halos surviving the supernova shock, as illustrated in Figs.~\ref{fig:sfbox0} and \ref{fig:sfbox1}. Metals from the supernova are mixed into these structures, which collapse relatively soon thereafter. However, we do not find an enhanced [\element{Fe}/\element{H}]\ abundance among the stars formed after short recollapse times. We also see that it is difficult to predict which halos will have short recollapse times based purely on their dense gas mass at the moment that the Pop III supernova explodes: there is evidence for some degree of anticorrelation between $M_{\rm dense}$ and $t_{\rm rec}$, but the halos with the shortest recollapse times have similar dense gas masses to some halos with $t_{\rm rec} \sim \qty{100}{\mega\yr}$. This suggests that not just the amount of dense gas is important but also the details of how it is distributed relative to the supernova.

The median $[\element{Fe}/\element{H}]_\textrm{2nd}$ abundance of stars forming in the aftermath of low energy PISN is $-5.5$ which is \qty{-2.9}{\dex} smaller than in the high energy PISN case.
There is a positive correlation between $[\element{Fe}/\element{H}]_\textrm{2nd}$ and $[\element{Fe}/\element{H}]_\textrm{halo}$ with $r=0.59$ that increases to $r=0.88$ if both datasets of both supernova energies are combined.
This is in good agreement with the correlations of $ M_\textrm{vir}$, $ M_\textrm{b}$ or $ t_\textrm{rec}$ in the high energy PISN case.
We conclude that knowing the halo metallicity is sufficient to infer the stellar metallicity regardless of the halo mass.
On the other hand, the recollapse time is not a good indicator for $[\element{Fe}/\element{H}]_\textrm{2nd}$. 

In the halo with id 11, we observe that two first-generation stars form. These two stars are not a binary system, because their distance of closest approach during their lifetimes is \qty{3.9}{\pc} and
the total energy of the two-body system is mostly positive except for a short amount of time during the closest encounter, i.e.\ they are not gravitationally bound to each other. Rather, this appears to be a case where the stars are forming in separate dense structures that just happen to assemble at a very similar time. The two stars form and explode with a delay of \qty{13.8}{\kilo\yr}, effectively depositing a total energy of \qty{e52}{\erg} into their host halo. Despite this, this halo belongs to the group of quickly recollapsing halos having a recollapse time of just \qty{0.5}{\mega\yr}.
The total metal yield injected into this halo is larger by \qty{0.3}{\dex}. As a consequence, the $[\element{Fe}/\element{H}]_\textrm{halo}$ abundance of this halo at recollapse is largest. However, in other respects this halo is not an outlier.

To illustrate the non-homogeneous and non-isotropic nature of the gas at recollapse, we show projections of the star-forming regions at recollapse in Figs.~\ref{fig:sfbox0} and \ref{fig:sfbox1}.
The projections are along the z-axis centred at the locations of star formation of a comoving \qty{2}{\kilo\pc\per\hubble} wide bounding box.
The projected metallicity can reach $[\element{Fe}/\element{H}] \sim -4.5$ but there are significant spatial variations.
Additionally, we show projections of the innermost star-forming region in a \qty{20}{\pc} wide bounding box as insets.

Finally, we note that all the second generation stars forming in our simulation do so in halos that previously hosted Pop III star formation, i.e.\ they are all examples of what is commonly termed internal enrichment. We see no examples of stars forming in externally enriched halos during the period simulated, although we cannot rule out this occurring at much later times.

\section{Discussion} \label{sec:discussion}

\subsection{Comparison with Observations} \label{sec:discussion:observations}
There have been claims of observational evidence for stars that are descendants of Pop III PISNe \citep{Aoki+2014,Aguado+2023,Vanzella+2023,Xing+2023,Ji+2024}, but these claims have typically not held up to further observational investigation
\citep{Koutsouridou+2024,Skuladottir+2024a,Bonifacio+2025}. 
Absorption system that are enriched by Pop III stars, but have not collapsed and formed stars yet, serve as another indirect method to constrain the properties of Pop III stars \citep{DOdorico+2013,Saccardi+2023,Salvadori+2023,Vanni+2024}.
Direct observations of high redshift PISNe are possible with JWST, but the predicted event rates are small \citep[see the discussion in][]{KlessenGlover2023} although there has recently been an observational claim of a \element{He}\textsc{ii} emitter as a potential Pop III candidate \citep{Maiolino+2026}.
We therefore investigate here whether our results agree with general observational trends rather than asking whether they can reproduce the abundances observed in individual objects.

\begin{figure}
	\centering
	\includegraphics[width=\linewidth]{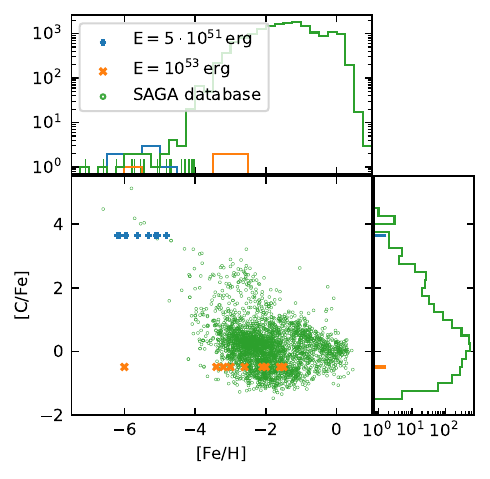}
	\caption{
		Metal distribution function (MDF) $[\element{Fe}/\element{H}]$ and carbon distribution function (CDF) $[\element{C}/\element{Fe}]$ joint (bottom left) and marginalised distributions (top, right, respectively) of second-generation stars in this simulation (blue) and in \citet[orange]{Magg+2022}.
		The distributions are distinct except for statistical outliers.
		For comparison, the MDF and CDF derived from the SAGA database \citep[data retrieved on 27 Feb 2026]{Suda+2017} is shown (green).
	}
	\label{fig:mdf}
\end{figure}

Since all the second-generation stars in our simulation are enriched by the same type of supernova with identical yields, they by definition all have the same relative abundances of heavy elements, with only the total metallicity (quantified here by [\element{Fe}/\element{H}]) varying from star to star. Specifically, the second-generation stars in our model have $[\element{C}/\element{Fe}]\sim 3.6$ \citep{Nomoto+2013}, in contrast to those in the high-energy PISN simulation, which had $[\element{C}/\element{Fe}] = -0.4$ \citep{HegerWoosley2002}. In Fig.~\ref{fig:mdf}, we show the metallicity distribution function (MDF) and carbon-to-iron ratio for the stars formed in our simulation and in the \citet{Magg+2022} simulation. For comparison, we also show the distributions functions for stars from the SAGA database \citep[data retrieved on 27 Feb 2026]{Suda+2017}. This collects the most metal poor stars observed in the Milky Way or (dwarf) galaxies in the Local Group.

The most metal poor star in the SAGA database has $[\element{Fe}/\element{H}]=-7.1$, which is smaller than the $[\element{Fe}/\element{H}]$ abundance of any star in our simulation.
However, our simulation results are broadly compatible with the observed metallicity and $[\element{C}/\element{Fe}]$ abundance range of the stars in the SAGA database, except for the single low metallicity outlier in the \citet{Magg+2022} simulation. The second generation stars formed in our simulation have a much lower median metallicity ($[\element{Fe}/\element{H}]\sim -5.5$) than those formed in the \citet{Magg+2022} simulation ($[\element{Fe}/\element{H}]\sim -2.6$), while having a much higher [\element{C}/\element{Fe}]\ abundance, but this agrees well with the trend towards higher [\element{C}/\element{Fe}]\ at lower [\element{Fe}/\element{H}]\ seen in the observational data.

Therefore, second generation stars enriched by a single PISNe should be found in the same parts of the [\element{C}/\element{Fe}]--[\element{Fe}/\element{H}]\ observational plane as the EMP stars observed in the halo, regardless of whether the PISNe has high or low energy. The fact that PISNe descendants are not observed in this region is therefore strong evidence that such stars must actually have been very rare.

\subsection{Comparison with Semi-analytic Models} \label{sec:discussion:sam}

Semi-analytic models (SAMs) demonstrate their strength in modelling large ensembles that easily allow the variation of statistical parameters such as the IMF.
The SAMs that we consider here solve an analytic model on top of results of cosmological N-body simulations.
The parameters are calibrated in such a way that the SAM correctly predicts statistical distributions or quantities observed in the local Universe.
A direct comparison with our case study is more difficult because the temporal extent of our simulation is limited.

One of the results of the A-SLOTH SAM \citep{Hartwig+2022,Magg+2022b} explored in \citet{Hartwig+2024} is the distribution of recovery times. Their definition of this quantity is the time between the supernova explosion and the next star-formation event within the same main progenitor branch of the dark matter merger tree. Thus, it is identical to the way 
the recollapse time is defined in this paper, allowing us to directly compare the results. Their distribution of recovery times ranges from $\ll \qty{1}{\mega\yr}$ to $> \qty{100}{\mega\yr}$, but has a clear peak at about \qty{30}{\mega\yr}.
Our recovery times are somewhat shorter, with a median value of around \qty{10}{\mega\yr}. One reason for this difference may be the different values assumed for the PISN explosion energy: A-SLOTH assumes that all PISNe explode with an energy of \qty{3.3e52}{\erg} \citep[see Fig. 2]{Hartwig+2022}, almost an order of magnitude larger than the value adopted in our study. In addition, the A-SLOTH model cannot account for the influence of dense sub-structure within the halos, which as we have seen appears to shorten $t_{\rm rec}$.
Note also that we have shown that the recollapse time does not strongly correlate with any other quantity except the mass of the halo. More massive halos reaccrete mass faster but metals and gas pushed by stronger supernova shocks requires more time to reaccrete and recollapse.

\citet{Koutsouridou+2023} use the NEFERTITI SAM to study the location in the [\element{C}/\element{Fe}] --[\element{Fe}/\element{H}]\ plane of stars enriched by CCSNe of various energies, hypernovae and PISN \citep[see also][]{Koutsouridou+2026}. They find that stars enriched by PISNe have a large range of different metallicities, from values as small as $[\element{Fe}/\element{H}] \sim -7$ to ones as large as $[\element{Fe}/\element{H}] \sim -2$, in good agreement with the combined results of our study and that of \citet{Magg+2022}. They also find a pronounced trend of increasing [\element{C}/\element{Fe}]\ with decreasing [\element{Fe}/\element{H}], with values that largely agree with those in our study, although we find a greater degree of scatter at very low metallicites than in their model. 
In a more recent study, \citet{Koutsouridou+2026} present results showing how the recovery time varies as a function of the supernova explosion energy. For the energy adopted here, $E\le\qty{5e51}{\erg}$, they find values between 1 and 100~Myr, with a mean value of 5~Myr, although noting this likely being a lower limit due to the implementation of the SN feedback.

In our simulations, we find that at the time of recollapse, there is a strong correlation between the halo-averaged and stellar [\element{Fe}/\element{H}]\ abundance with a correlation coefficient $r=0.59$ that increases to $r=0.88$ when we combine our results with those from by \citet{Magg+2022}. This correlation is suggestive of rapid and efficient but not perfect mixing of the metal yields.
In contrast, most SAMs assume instantaneous mixing of the halo or galaxy gas, although a notable exception is the A-SLOTH model, which implements a metallicity shift of the newly-formed stars with respect to the metallicity of the host galaxy, following \citet{Tarumi+2020}. Other SAMs are restricted to implement a successive \citep{Hartwig+2022,Ventura+2024,Koutsouridou+2026} or effective \citep{Salvadori+2019} metal mixing enrichment of the intergalactic medium (IGM).
We do not probe the metal enrichment of the IGM directly rendering a direct comparison difficult, but we emphasise that SAMs gain accuracy by implementing the expansion of the Sedov-Taylor blast wave that controls metal enrichment.

\subsection{Comparison with Hydrodynamic Simulations} \label{sec:discussion:hydro}

Hydrodynamic simulations are more expensive to carry out than SAMs, but unlike SAMs they have the advantage that they can be used to directly follow the mixing of supernova ejecta into the surrounding gas, rather than requiring one to make assumptions about the effectiveness of this process. Many numerical studies of the production and dispersal of metals at high redshift have been carried out, but the majority either focus on a single Pop III supernova \citep[e.g.][]{Bromm+2003,Karlsson+2008,Greif+2010,Whalen+2013,Chiaki+2018} or study enrichment driven by stars forming in much larger galaxies \citep[e.g.][]{Johnson+2013,OShea+2015,Nelson+2019}. 
We therefore restrict our comparison to \citet{Ritter+2015,Ritter+2016}, \citet{LatifSchleicher2020}, \citet{Brauer+2025a}, and \citet{Mead+2025a} who follow a similar approach to the one used in this paper and by \citet{Magg+2022} and provide a larger statistical sample.

The simulations by \citet{Ritter+2015,Ritter+2016} are cosmological zoom-in simulations of a comparable volume to our simulation, but they only resolve the most massive $\sim \qty{e6}{\Msun}$ halo in the simulated volume. They carry out three variants: two in which they follow the explosion of a single massive star \citetext{\citealp{Ritter+2015}, \textsc{1sn}; \citealp{Ritter+2016}} and a third in which they look at the impact of a small cluster of massive stars \citep[\textsc{7sn}]{Ritter+2015}. All stars explode as CCSNe, each depositing \qty{e51}{\erg} of energy. 
The dense recollapsing gas clouds have $Z\sim \num{6e-6}$ and $Z\sim \num{5e-7}$ in the \textsc{1sn} and \textsc{7sn} models, respectively \citep{Ritter+2015}.
Adopting the CCSN yields from \citet{HegerWoosley2010}, we can convert their quoted values for the total metallicity to an $[\element{Fe}/\element{H}] \lesssim -3.6$, which is larger by up to $\sim \qty{2}{\dex}$ than in our simulation.
In \citet{Ritter+2016}, a dense gas cloud is superficially enriched by the supernova ejecta of a nearby supernova and the ejecta mix into the cloud reaching \qtyrange{e-4}{e-3}{\Zsun}.
This is a similar process to that we find with the halos with ids 2 and 11, but the metallicity $Z$ of second generation stars in \citet{Ritter+2016} is larger by \qtyrange{0.2}{0.5}{\dex} using Eq. \eqref{eq:total_metallicity} and $\unit{\Zsun}=0.02$.
As long as no specific element abundances are taken into account, the total metallicity of second-generation stars forming after CCSNe is not significantly different from stars forming after a low energy PISNe.

A closer comparison is possible with the simulation by \citet{LatifSchleicher2020} who study metal enrichment by PISNe.
They perform a cosmological zoom-in simulation of a comoving volume of \qty{1}{\cubic\mega\pc\per\cubic\hubble}, zooming into five halos.
The halos host stars of \qty{182}{\Msun} exploding with an energy of \qty{3.6e52}{\erg} in a gas that is constantly heated by a Lyman-Werner background radiation with $J_{21}=\qty{e-21}{\erg\per\square\centi\metre\per\second\per\hertz\per\steradian}$.
Three of their five star-forming halos are evaporated by the supernova explosion.
In the other two halos, dense gas surviving the radiative feedback of the first stars and the supernova shock is able to form stars within \qty{2}{\mega\yr} after the supernova, similar to the very short recollapse times in our simulation.
The other halos require about \qty{31}{\mega\yr} until second-generation stars form which is similar to \citet{Magg+2022}.
The metallicity of Pop II stars peaks at $Z\gtrsim \num{e-4}$ which is at $[\element{Fe}/\element{H}]\sim -5.8$ adopting the yields by \citet{HegerWoosley2002}.
The longer recollapse time may be due to a combination of a higher explosion energy and the external Lyman-Werner radiation that heats the gas.
The latter could explain a significantly lower metallicity if warmer gas more efficiently mixes the metals into the ISM and inter-halo medium.

The AEOS simulation, first presented by \citet{Brauer+2025a}, aims to model the formation of the first galaxies with star-by-star resolution to trace metal enrichment \citep{Mead+2025a}, similar to the approach used in our simulation.  Critically, the AEOS simulation adopts an IMF in the mass range \qtyrange{1}{100}{\Msun} with a characteristic mass \qty{10}{\Msun} which means, the simulation exclusively hosts stars that explode as CCSN with \qty{e51}{\erg}.
Their Pop II stars, i.e. stars forming in gas with $Z>\qty{e-5}{Z_\odot}\sim \num{e-7}$, form in gas with $[\element{Fe}/\element{H}]=-5.1$ and $[\element{C}/\element{Fe}]=1$ (best-fit).
Even these low energy supernovae are capable of ejecting large fractions, if not all, metals out of the halo.
\citet{Mead+2025b} find a mixing timescale $\tau_\textrm{chem}\sim\qty{7}{\mega\yr}$ until the abundances produced by individual stars have diluted in the ISM to an IMF-averaged abundance pattern.
Due to the fact that we do not employ an IMF and resolve metal enrichment only by tracers, we cannot directly compare this timescale to our simulation results.
The median recollapse time in our simulation is \qty{9.5}{\mega\yr} and until this time, the chemical composition remains the one enriched by a single PISN.
The AEOS20 variant \citep{Brauer+2025b} increases the characteristic mass to \qty{20}{\Msun} and the mass range of Pop III stars to \qty{300}{\Msun}, but the stars still explode as CCSNe.
This leads to an enhanced odd-even pattern and stronger $\alpha$-element enhancement of the metal-enriched gas.

\subsection{Implications for Pop III Properties}

Taken together, our results and those of \citet{Magg+2022} demonstrate that if some Pop III stars explode as PISNe, this must be a relatively rare event, or else we would see evidence for it in observations of EMP stars. Although Pop III PISNe are capable of producing large amounts of metals, the majority of their descendants have metallicities placing them within the  
range observed in EMP stars. There is, however, no spectroscopic confirmation of PISN enrichment in these stars, implying that the vast majority were enriched by CCSNe, which we propose as a favoured model. In this context, we highlight recent theoretical work by \citet{ShardaMenon2025} and \citet{Sharda+2025}, who find in radiation-magnetohydrodynamic simulations of Pop III star formation that magnetic fields keep the gas temperature low which enhances fragmentation of the gas cloud leading to smaller protostar masses.

The internal physics of CCSNe is still not very well constrained despite considerable effort \citetext{c.f. \citealp{Janka2025} for a recent review}. Previous computations of Pop III supernovae left the supernova explosion energy as a free parameter, thus spanning up to \qty{2}{\dex} \citep{HegerWoosley2010,LimongiChieffi2012}.
Better constraining the CCSN explosion energy is of high relevance because the binding energy of the first star-forming halos is of a similar order of magnitude as the explosion energy, and we found it to be a predictor for the recollapse time.
As a direct consequence, a faster recollapse time implies more internal recycling of supernova ejecta, stronger mixing in place and a shortened time until the abundance pattern reaches the population average \citep[c.f.][]{Brauer+2025b}.

Other observational methods apart from stellar archaeology may also turn out to be fruitful for constraining the properties of Pop III SNe in the near future.
Since supermassive Pop III stars may reach several hundreds of \unit{\Msun}, the gravitational collapse that occurs once the core becomes susceptible to the general relativity instability may produce bright light curves.
These could be observable with JWST, the \textit{Roman Space Telescope} (RST) or \textit{Euclid} \citep{Jockel+2026}.
A possible explanation for the observation of the Little Red Dots (LRDs) with JWST is a transient event.
During the PISN, the light curve could be bright enough to be observable with JWST \citep{Ferrara+2026,Jeon+2026}.
Finally, the influence of Pop III stars on the global 21-cm signal, which will soon be probed with observations by SKA, may shine new light on the  population statistics of these stars \citep{Gessey-Jones+2025}.

For future theoretical work, there is a strong requirement for larger statistics.
On the theoretical modelling side, hydrodynamic simulations progressed from isolated halos with individual supernovae to embeddings into small cosmological volumes.
The next transition requires embedding into larger cosmological contexts with about an order of magnitude larger samples.
The physical models mostly converged on suitable parameter sets (e.g. the sink/star formation threshold).
Radiative transfer is usually included but whether magnetic fields or cosmic rays play a vital role in determining the physics of star formation, especially in weakly metal-enriched gas, or whether they blur under the stochasticity of the collapse of molecular clouds remains to be explored.
From a more general perspective, we should find way to disentangle stellar generations, the historic Population concept and relevant timescale to quantify chemical mixing and renewed star formation.
This assists in making results comparable across simulations, SAMs and observations.

\subsection{Shortcomings of the Adopted Model} \label{sec:discussion:shortcomings}
There are several caveats related to our simulation which should be borne in mind when interpreting our results. 

First, although the goal of this simulation was to investigate the impact of the least energetic PISNe, we note that the explosion energy adopted in this work may differ from the true value. The lower mass limit of \qty{140}{\Msun} for PISNe was proposed by \citet{HegerWoosley2002} and \citet{Chen+2014}, but it overlaps with the upper limit of pulsational PISN \citep{Leung+2019}, where the star undergoes several strong pulses ejecting its upper layers before it goes supernova. The more recent computation by \citet{Costa+2025} puts the lower mass limit of PISNe at \qty{130}{\Msun} (\qty{110}{\Msun} for pulsational PISNe), although these authors do not comment on the explosion energy associated with the lowest mass PISNe. On the other hand, \citet{Takahashi+2018} find a larger minimum PISNe mass: \qty{180}{\Msun} in the non-rotating case, with an explosion energy of approximately \qty{1.7e52}{\erg}, or \qty{160}{\Msun} in the rotating case, with an explosion energy of between $7 \times 10^{51}$ and $9 \times 10^{51} \: {\rm erg}$. This could imply that all real PISNe are more energetic than those whose impact we have attempted to model here.

Second, we note that although we have adopted yields for a \qty{140}{\Msun} PISN from \citet{Nomoto+2013}, other predictions exist which may result in the production of very different amounts of iron. To take an extreme example, the iron production factor $f_\textrm{Fe} = M_\textrm{Fe} / M_\textrm{ZAMS}$ of the \qty{140}{\Msun} PISN model from \citet{HegerWoosley2002} is only $f_\textrm{Fe}=10^{-14.7}$, many orders of magnitude that the \citet{Nomoto+2013} value ($f_{\rm Fe} \simeq 3.6 \times 10^{-3}$). The lowest energy PISN of \citet{Costa+2025} produces $f_\textrm{Fe} \sim 10^{-5}$, while the values from \citet{Takahashi+2018} range between $f_\textrm{Fe} \sim 10^{-4}$ and a value similar to \citet{Nomoto+2013} depending on the details of the stellar model. There is thus a considerable uncertainty involved in the amount of iron produced by the least energetic PISNe. Note, however, that our choice to use the \citet{Nomoto+2013} value for the yield is conservative in the sense that it amongst the largest available. Adopting a smaller value would shift the second generation stars formed in our simulation to lower [\element{Fe}/\element{H}], strengthening our conclusions.

Third, as previously noted, in our treatment of radiative feedback we adopt ionising photon production rates that are the same as used by \citet{Magg+2022} and hence inconsistent with the values that we would expect \qty{140}{\Msun} Pop III stars to actually produce. We chose to do this in order to minimize the number of differences between our simulation and \citet{Magg+2022}, allowing us to focus on the impact of the supernova energy, but as a consequence the radiation feedback in our simulation is somewhat more effective than should actually be the case. 

Fourth, our model is agnostic to metals, and we follow metal enrichment only with tracer particles such that quantities can easily be derived from Eq.~\eqref{eq:densitysquared} as we demonstrate with the total metallicity $\log Z$ in Eq. \eqref{eq:total_metallicity}. This means that we do not account for any impact of metal-line cooling. At the densities and metallicities studied in this simulation, this should be a good approximation \citep{Jappsen+2007}, but it would not remain so if we were to follow the collapse of the metal-enriched gas up to

Finally, we want to stress that parts of this simulation were designed to serve as a numerical experiment. 
Some parameter choices we deliberately modified, others were chosen to be consistent with \citet{Magg+2022} for a fair comparison.
Future work will aim to explore different parameters in more realistic configuration and using current state-of-the-art knowledge about the first stars \citep[e.g. with the subgrid models presented in ][]{Liu+2024,Gurian+2026}.
\section{Conclusion} \label{sec:conclusion}

We investigated the hypothesis that very massive Pop III stars in the mass range \qtyrange{140}{270}{\Msun}, which end their lives as pair-instability supernovae releasing large amounts of energy and producing substantial amounts of metals, could enrich their host halos to metallicities much greater than $[\element{Fe}/\element{H}] = -3$ \citep{Salvadori+2019}. 
Such enrichment would imply that second-generation stars formed from gas enriched by PISN would have metallicities larger than the value used to define EMP stars, thereby explaining why stars with unambiguous signatures of PISN enrichment are not found in surveys of EMP stars. Our investigation complements an earlier study by \citet{Magg+2022}, who showed that although PISN with masses and explosion energies at the upper limit for such supernovae can in some cases produce regions enriched to $[\element{Fe}/\element{H}] > -3$, many of the stars forming in gas enriched by these SNe have $[\element{Fe}/\element{H}] < -3$ and hence would be classified as EMP stars. In our study, we examined PISN at the other end of the allowed mass range, which have explosion energies around 20 times smaller, to see whether a similar result also holds in that case.

To do this, we performed a cosmological hydrodynamical simulation that followed the formation of Pop III stars starting from initial conditions through the collapse of their minihalos. For the purposes of this numerical experiment, we assumed that each star-forming minihalo produced exactly one low mass PISN with $M = (\qty{140}{\Msun})$ and explosion energy $E = \qty{5e51}{\erg}$. We followed the simulation for approximately 50~Myr after the first PISN explosion, allowing us to study the aftermath of multiple enrichment events within the simulation volume.

We find that in all cases, the second generation stars that form in the simulation do so in halos that previously hosted a Pop III SN, i.e.\ we see no cases of external enrichment of nearby halos. On average, the second-generation stars form with low metallicities, $[\element{Fe}/\element{H}]\sim -5.5$, and the halo average [\element{Fe}/\element{H}]\ strongly correlates with the [\element{Fe}/\element{H}]\ abundance of the second-generations stars.
While the theoretically derived iron yield of the low-energy PISNe is \qty{4}{\dex} lower than in the high-energy case, the [\element{Fe}/\element{H}]\ abundance of second-generation stars in a simulation-to-simulation comparison exhibits on average an [\element{Fe}/\element{H}]\ abundance reduced by only \qty{2.9}{\dex} \citep{Magg+2022}. This discrepancy arises because of the difference in the explosion energy which 
limits the ability of the ejecta to escape from the halo's potential well.
Metals therefore remain close to the explosion site rather than flowing out of the halo to mix and dilute into the inter-halo gas.
Consequently, the reduction in the metallicity is less than one would expect based purely on the SN yield, as the SN ejecta are mixed into a much smaller volume of gas.

The distributions of [\element{C}/\element{Fe}]\ and [\element{Fe}/\element{H}]\ produced by the low energy PISN studied here and the high energy PISN studied in \citet{Magg+2022}
overlap with the observed abundances of EMP stars \citep[SAGA database]{Suda+2017}.
However, none of the observed stars in that database feature the pronounced odd-even abundance pattern that would render them as distinctive Pop III PISN descendants \citep{HegerWoosley2002}.
There is hence little or no evidence that EMP stars resemble mono-enriched descendants of Pop III which exploded as a PISN.

Given these results, we conclude that alternative enrichment channels dominate the early chemical evolution of the Universe. If any Pop III PISNe occurred, they must have been rare events and the bulk of the enrichment produced in the observed EMP stars must have come from lower mass CCSNe and/or hypernovae. We plan to explore how the feedback of these supernovae shape the chemistry of second-generation stars in future work.

\begin{acknowledgement}
We are grateful to Mattis Magg for sharing data, code and providing valuable comments. We thank Philipp Girichdis, Boyuan Liu and Bipradeep Saha for useful discussions. The authors acknowledge financial support from the ERC via Synergy Grant ``ECOGAL'' (project ID 855130) and from the German Excellence Strategy via the Heidelberg Cluster ``STRUCTURES'' (EXC 2181 - 390900948). In addition AK acknowledges funding by the Cusanuswerk as part of the German scholarship system funded by the BMFTR. RSK is grateful for funding from the German Ministry for Economy and Energy (BMWE) in project ``MAINN'' (funding ID 50OO2206), and from DFG and ANR for project ``STARCLUSTERS'' (funding ID KL 1358/22-1). 
The authors also acknowledge support by the State of Baden-Württemberg through bwHPC, bwVisu and DFG through grant INST 35/1597-1 FUGG.
The authors gratefully acknowledge the data storage service SDS@hd supported by the State of Baden-Württemberg and DFG through grant INST 35/1503-1 FUGG.
Software used: \textsc{numpy} \citep{Harris+2020}, \textsc{scipy} \citep{Virtanen+2020}, \textsc{matplotlib}\citep{Hunter2007}, \textsc{yt} \citep{Turk+2011}, and \textsc{scida} \citep{ByrohlNelson2024}.
\end{acknowledgement}

\bibliographystyle{aa} 
\bibliography{literature} 

\begin{thebibliography}{96}
\expandafter\ifx\csname natexlab\endcsname\relax\def\natexlab#1{#1}\fi

\bibitem[{Aguado {et~al.}(2023)Aguado, Salvadori, Sk{\'u}lad{\'o}ttir, Caffau,
  Bonifacio, Vanni, Gelli, Koutsouridou, \& Amarsi}]{Aguado+2023}
Aguado, D.~S., Salvadori, S., Sk{\'u}lad{\'o}ttir, {\'A}., {et~al.} 2023,
  \mnras, 520, 866

\bibitem[{Aoki {et~al.}(2014)Aoki, Tominaga, Beers, Honda, \& Lee}]{Aoki+2014}
Aoki, W., Tominaga, N., Beers, T.~C., Honda, S., \& Lee, Y.~S. 2014, Science,
  345, 912

\bibitem[{Asplund {et~al.}(2009)Asplund, Grevesse, Sauval, \&
  Scott}]{Asplund+2009}
Asplund, M., Grevesse, N., Sauval, A.~J., \& Scott, P. 2009, \araa, 47, 481

\bibitem[{Beers {et~al.}(2005)Beers, Christlieb, Norris, Bessell, Wilhelm,
  Allende~Prieto, Yanny, Rockosi, Newberg, Rossi, \& Lee}]{Beers+2005}
Beers, T.~C., Christlieb, N., Norris, J.~E., {et~al.} 2005, in From {{Lithium}}
  to {{Uranium}}: {{Elemental Tracers}} of {{Early Cosmic Evolution}}, Vol.
  228, eprint: arXiv:astro-ph/0508423, 175--183

\bibitem[{Behroozi {et~al.}(2013{\natexlab{a}})Behroozi, Wechsler, \&
  Wu}]{Behroozi+2013}
Behroozi, P.~S., Wechsler, R.~H., \& Wu, H.-Y. 2013{\natexlab{a}}, \apj, 762,
  109

\bibitem[{Behroozi {et~al.}(2013{\natexlab{b}})Behroozi, Wechsler, Wu, Busha,
  Klypin, \& Primack}]{Behroozi+2013a}
Behroozi, P.~S., Wechsler, R.~H., Wu, H.-Y., {et~al.} 2013{\natexlab{b}}, \apj,
  763, 18

\bibitem[{Bonifacio {et~al.}(2025)Bonifacio, Caffau, Fran{\c c}ois, \&
  Spite}]{Bonifacio+2025}
Bonifacio, P., Caffau, E., Fran{\c c}ois, P., \& Spite, M. 2025, \aapr, 33, 2

\bibitem[{Brauer {et~al.}(2025{\natexlab{a}})Brauer, Emerick, Mead, Ji, Wise,
  Bryan, Mac~Low, C{\^o}t{\'e}, Andersson, \& Frebel}]{Brauer+2025a}
Brauer, K., Emerick, A., Mead, J., {et~al.} 2025{\natexlab{a}}, \apj, 980, 41

\bibitem[{Brauer {et~al.}(2025{\natexlab{b}})Brauer, Mead, Wise, Bryan, Low,
  Ji, Emerick, Andersson, Frebel, \& C{\^o}t{\'e}}]{Brauer+2025b}
Brauer, K., Mead, J., Wise, J.~H., {et~al.} 2025{\natexlab{b}}, \apj, 993, 2

\bibitem[{Bromm {et~al.}(2003)Bromm, Yoshida, \& Hernquist}]{Bromm+2003}
Bromm, V., Yoshida, N., \& Hernquist, L. 2003, \apj, 596, L135

\bibitem[{Byrohl \& Nelson(2024)}]{ByrohlNelson2024}
Byrohl, C. \& Nelson, D. 2024, The Journal of Open Source Software, 9, 6064

\bibitem[{Chen {et~al.}(2014)Chen, Heger, Woosley, Almgren, \&
  Whalen}]{Chen+2014}
Chen, K.-J., Heger, A., Woosley, S., Almgren, A., \& Whalen, D.~J. 2014, \apj,
  792, 44

\bibitem[{Chiaki {et~al.}(2018)Chiaki, Susa, \& Hirano}]{Chiaki+2018}
Chiaki, G., Susa, H., \& Hirano, S. 2018, \mnras, 475, 4378

\bibitem[{Clark {et~al.}(2011{\natexlab{a}})Clark, Glover, Klessen, \&
  Bromm}]{Clark+2011a}
Clark, P.~C., Glover, S. C.~O., Klessen, R.~S., \& Bromm, V.
  2011{\natexlab{a}}, \apj, 727, 110

\bibitem[{Clark {et~al.}(2011{\natexlab{b}})Clark, Glover, Smith, Greif,
  Klessen, \& Bromm}]{Clark+2011}
Clark, P.~C., Glover, S. C.~O., Smith, R.~J., {et~al.} 2011{\natexlab{b}},
  Science, 331, 1040

\bibitem[{Costa {et~al.}(2025)Costa, Shepherd, Bressan, Addari, Chen, Fu,
  Volpato, Nguyen, Girardi, Marigo, Mazzi, Pastorelli, Trabucchi, Bossini, \&
  Zaggia}]{Costa+2025}
Costa, G., Shepherd, K.~G., Bressan, A., {et~al.} 2025, \aap, 694, A193

\bibitem[{{de Bennassuti} {et~al.}(2017){de Bennassuti}, Salvadori, Schneider,
  Valiante, \& Omukai}]{deBennassuti+2017}
{de Bennassuti}, M., Salvadori, S., Schneider, R., Valiante, R., \& Omukai, K.
  2017, \mnras, 465, 926

\bibitem[{D'Odorico {et~al.}(2013)D'Odorico, Cupani, Cristiani, Maiolino,
  Molaro, Nonino, Centuri{\'o}n, Cimatti, {di Serego Alighieri}, Fiore,
  Fontana, Gallerani, Giallongo, Mannucci, Marconi, Pentericci, Viel, \&
  Vladilo}]{DOdorico+2013}
D'Odorico, V., Cupani, G., Cristiani, S., {et~al.} 2013, \mnras, 435, 1198

\bibitem[{{Ferrara} {et~al.}(2026){Ferrara}, {Carniani}, {Morishita}, \&
  {Stiavelli}}]{Ferrara+2026}
{Ferrara}, A., {Carniani}, S., {Morishita}, T., \& {Stiavelli}, M. 2026, arXiv
  e-prints, arXiv:2601.07374

\bibitem[{Frebel \& Norris(2015)}]{FrebelNorris2015}
Frebel, A. \& Norris, J.~E. 2015, \araa, 53, 631

\bibitem[{Galli \& Palla(2013)}]{GalliPalla2013}
Galli, D. \& Palla, F. 2013, \araa, 51, 163

\bibitem[{Genel {et~al.}(2013)Genel, Vogelsberger, Nelson, Sijacki, Springel,
  \& Hernquist}]{Genel+2013}
Genel, S., Vogelsberger, M., Nelson, D., {et~al.} 2013, \mnras, 435, 1426

\bibitem[{{Gessey-Jones} {et~al.}(2025){Gessey-Jones}, Sartorio, Bevins,
  Fialkov, Handley, {de Lera Acedo}, Mirouh, Izzard, \&
  Barkana}]{Gessey-Jones+2025}
{Gessey-Jones}, T., Sartorio, N.~S., Bevins, H. T.~J., {et~al.} 2025, Nature
  Astronomy, 9, 1268

\bibitem[{Greif {et~al.}(2010)Greif, Glover, Bromm, \& Klessen}]{Greif+2010}
Greif, T.~H., Glover, S. C.~O., Bromm, V., \& Klessen, R.~S. 2010, \apj, 716,
  510

\bibitem[{Greif {et~al.}(2011)Greif, Springel, White, Glover, Clark, Smith,
  Klessen, \& Bromm}]{Greif+2011}
Greif, T.~H., Springel, V., White, S. D.~M., {et~al.} 2011, \apj, 737, 75

\bibitem[{{Gurian} {et~al.}(2026){Gurian}, {Liu}, {Jeong}, {Hosokawa},
  {Hirano}, {Bromm}, \& {Yoshida}}]{Gurian+2026}
{Gurian}, J., {Liu}, B., {Jeong}, D., {et~al.} 2026, arXiv e-prints,
  arXiv:2604.26006

\bibitem[{Hahn \& Abel(2011)}]{HahnAbel2011}
Hahn, O. \& Abel, T. 2011, \mnras, 415, 2101

\bibitem[{Harris {et~al.}(2020)Harris, Millman, {van der Walt}, Gommers,
  Virtanen, Cournapeau, Wieser, Taylor, Berg, Smith, Kern, Picus, Hoyer, {van
  Kerkwijk}, Brett, Haldane, {del R{\'i}o}, Wiebe, Peterson,
  {G{\'e}rard-Marchant}, Sheppard, Reddy, Weckesser, Abbasi, Gohlke, \&
  Oliphant}]{Harris+2020}
Harris, C.~R., Millman, K.~J., {van der Walt}, S.~J., {et~al.} 2020, Nature,
  585, 357

\bibitem[{Hartwig {et~al.}(2024)Hartwig, Lipatova, Glover, \&
  Klessen}]{Hartwig+2024}
Hartwig, T., Lipatova, V., Glover, S. C.~O., \& Klessen, R.~S. 2024, \mnras,
  535, 516

\bibitem[{Hartwig {et~al.}(2022)Hartwig, Magg, Chen, Tarumi, Bromm, Glover, Ji,
  Klessen, Latif, Volonteri, \& Yoshida}]{Hartwig+2022}
Hartwig, T., Magg, M., Chen, L.-H., {et~al.} 2022, \apj, 936, 45

\bibitem[{Heger \& Woosley(2002)}]{HegerWoosley2002}
Heger, A. \& Woosley, S.~E. 2002, \apj, 567, 532

\bibitem[{Heger \& Woosley(2010)}]{HegerWoosley2010}
Heger, A. \& Woosley, S.~E. 2010, \apj, 724, 341

\bibitem[{Hirano {et~al.}(2015)Hirano, Hosokawa, Yoshida, Omukai, \&
  Yorke}]{Hirano+2015}
Hirano, S., Hosokawa, T., Yoshida, N., Omukai, K., \& Yorke, H.~W. 2015,
  \mnras, 448, 568

\bibitem[{Hunter(2007)}]{Hunter2007}
Hunter, J.~D. 2007, Computing in Science and Engineering, 9, 90

\bibitem[{Janka(2025)}]{Janka2025}
Janka, H.-T. 2025, Annual Review of Nuclear and Particle Science, 75, 425

\bibitem[{Jappsen {et~al.}(2007)Jappsen, Glover, Klessen, \&
  Mac~Low}]{Jappsen+2007}
Jappsen, A.~K., Glover, S. C.~O., Klessen, R.~S., \& Mac~Low, M.~M. 2007, \apj,
  660, 1332

\bibitem[{Jaura {et~al.}(2018)Jaura, Glover, Klessen, \&
  Paardekooper}]{Jaura+2018}
Jaura, O., Glover, S. C.~O., Klessen, R.~S., \& Paardekooper, J.~P. 2018,
  \mnras, 475, 2822

\bibitem[{Jaura {et~al.}(2020)Jaura, Magg, Glover, \& Klessen}]{Jaura+2020}
Jaura, O., Magg, M., Glover, S. C.~O., \& Klessen, R.~S. 2020, \mnras, 499,
  3594

\bibitem[{{Jeon} {et~al.}(2026){Jeon}, {Bromm}, {Venditti}, {Finkelstein}, \&
  {Hsiao}}]{Jeon+2026}
{Jeon}, J., {Bromm}, V., {Venditti}, A., {Finkelstein}, S.~L., \& {Hsiao}, T.
  Y.-Y. 2026, \apj, 1001, 3

\bibitem[{Ji {et~al.}(2024)Ji, Curtis, Storm, Chandra, Schlaufman, Stassun,
  Heger, Pignatari, {Price-Whelan}, Bergemann, Stringfellow, Fr{\"o}hlich,
  Reggiani, Holmbeck, Tayar, Shah, Griffith, Laporte, Casey, Hawkins, Horta,
  Cerny, Thibodeaux, Usman, Amarante, Beaton, Cargile, Chiappini, Conroy,
  Johnson, Kollmeier, Li, Loebman, Meynet, Bizyaev, Brownstein, Gupta,
  Morrison, Pan, Ramirez, Rix, \& {S{\'a}nchez-Gallego}}]{Ji+2024}
Ji, A.~P., Curtis, S., Storm, N., {et~al.} 2024, \apj, 961, L41

\bibitem[{Jockel {et~al.}(2026)Jockel, Kawaguchi, Fujibayashi, \&
  Shibata}]{Jockel+2026}
Jockel, C., Kawaguchi, K., Fujibayashi, S., \& Shibata, M. 2026, \mnras, 545,
  staf1949

\bibitem[{Johnson {et~al.}(2013)Johnson, Dalla~Vecchia, \&
  Khochfar}]{Johnson+2013}
Johnson, J.~L., Dalla~Vecchia, C., \& Khochfar, S. 2013, \mnras, 428, 1857

\bibitem[{Karlsson {et~al.}(2008)Karlsson, Johnson, \& Bromm}]{Karlsson+2008}
Karlsson, T., Johnson, J.~L., \& Bromm, V. 2008, \apj, 679, 6

\bibitem[{Klessen \& Glover(2023)}]{KlessenGlover2023}
Klessen, R.~S. \& Glover, S. C.~O. 2023, \araa, 61, 65

\bibitem[{Koutsouridou {et~al.}(2024)Koutsouridou, Salvadori, \&
  Sk{\'u}lad{\'o}ttir}]{Koutsouridou+2024}
Koutsouridou, I., Salvadori, S., \& Sk{\'u}lad{\'o}ttir, {\'A}. 2024, \apj,
  962, L26

\bibitem[{{Koutsouridou} {et~al.}(2026){Koutsouridou}, {Salvadori},
  {Sk{\'u}lad{\'o}ttir}, {Gelli}, {Rusta}, {Querci}, {Aguado}, \&
  {Mori}}]{Koutsouridou+2026}
{Koutsouridou}, I., {Salvadori}, S., {Sk{\'u}lad{\'o}ttir}, {\'A}., {et~al.}
  2026, arXiv e-prints, arXiv:2605.00990

\bibitem[{Koutsouridou {et~al.}(2023)Koutsouridou, Salvadori,
  Sk{\'u}lad{\'o}ttir, Rossi, Vanni, \& Pagnini}]{Koutsouridou+2023}
Koutsouridou, I., Salvadori, S., Sk{\'u}lad{\'o}ttir, {\'A}., {et~al.} 2023,
  \mnras, 525, 190

\bibitem[{Latif \& Schleicher(2020)}]{LatifSchleicher2020}
Latif, M.~A. \& Schleicher, D. 2020, \apj, 902, L31

\bibitem[{Leung {et~al.}(2019)Leung, Nomoto, \& Blinnikov}]{Leung+2019}
Leung, S.-C., Nomoto, K., \& Blinnikov, S. 2019, \apj, 887, 72

\bibitem[{Limongi \& Chieffi(2012)}]{LimongiChieffi2012}
Limongi, M. \& Chieffi, A. 2012, \apjs, 199, 38

\bibitem[{Liu {et~al.}(2024)Liu, Gurian, Inayoshi, Hirano, Hosokawa, Bromm, \&
  Yoshida}]{Liu+2024}
Liu, B., Gurian, J., Inayoshi, K., {et~al.} 2024, \mnras, 534, 290

\bibitem[{Magg {et~al.}(2022{\natexlab{a}})Magg, Hartwig, Chen, \&
  Tarumi}]{Magg+2022b}
Magg, M., Hartwig, T., Chen, L.-H., \& Tarumi, Y. 2022{\natexlab{a}},
  Astrophysics Source Code Library, ascl:2209.001

\bibitem[{Magg {et~al.}(2022{\natexlab{b}})Magg, Schauer, Klessen, Glover,
  Tress, \& Jaura}]{Magg+2022}
Magg, M., Schauer, A. T.~P., Klessen, R.~S., {et~al.} 2022{\natexlab{b}}, \apj,
  929, 119

\bibitem[{Maiolino {et~al.}(2026)Maiolino, {\"U}bler, Perna, Witstok, Jones,
  {Perez-Gonzalez}, Nakajima, Rusta, Salvadori, Tacchella, Madau, Trussler,
  D'Eugenio, Ji, Scholtz, Carniani, Isobe, Katz, Arribas, Baker, B{\"o}ker,
  Bromm, Bunker, Charlot, Chevallard, Curti, {Curtis-Lake}, Eisenstein, Egami,
  Ferrara, Graziani, Hainline, Helton, Ivey, Jonson, Koller, Kumari, Marconi,
  Mazzolari, Laporte, Parlanti, Pascalau, Pentericci, Rinaldi, Robertson,
  Rodr{\'i}guez Del~Pino, Schneider, Venditti, Venturi, Willmer, Witten, \&
  Zamora}]{Maiolino+2026}
Maiolino, R., {\"U}bler, H., Perna, M., {et~al.} 2026, submitted;
  arXiv:2603.20362

\bibitem[{Mayer {et~al.}(2025)Mayer, Naab, Caselli, Ivlev, Grassi, Zier,
  Pakmor, Walch, \& Springel}]{Mayer+2025}
Mayer, A.~C., Naab, T., Caselli, P., {et~al.} 2025, \mnras, 543, 3321

\bibitem[{{Mead} {et~al.}(2025){Mead}, {Brauer}, {Bryan}, {Mac Low}, {Ji},
  {Wise}, {Andersson}, {Frebel}, {Emerick}, \& {C{\^o}t{\'e}}}]{Mead+2025b}
{Mead}, J., {Brauer}, K., {Bryan}, G.~L., {et~al.} 2025, arXiv e-prints,
  arXiv:2509.13580

\bibitem[{Mead {et~al.}(2025)Mead, Brauer, Bryan, Mac~Low, Ji, Wise, Emerick,
  Andersson, Frebel, \& C{\^o}t{\'e}}]{Mead+2025a}
Mead, J., Brauer, K., Bryan, G.~L., {et~al.} 2025, \apj, 980, 62

\bibitem[{Navarro {et~al.}(1997)Navarro, Frenk, \& White}]{Navarro+1997}
Navarro, J.~F., Frenk, C.~S., \& White, S. D.~M. 1997, \apj, 490, 493

\bibitem[{Nelson {et~al.}(2019)Nelson, Pillepich, Springel, Pakmor, Weinberger,
  Genel, Torrey, Vogelsberger, Marinacci, \& Hernquist}]{Nelson+2019}
Nelson, D., Pillepich, A., Springel, V., {et~al.} 2019, \mnras, 490, 3234

\bibitem[{Nomoto {et~al.}(2013)Nomoto, Kobayashi, \& Tominaga}]{Nomoto+2013}
Nomoto, K., Kobayashi, C., \& Tominaga, N. 2013, \araa, 51, 457

\bibitem[{O'Shea {et~al.}(2015)O'Shea, Wise, Xu, \& Norman}]{OShea+2015}
O'Shea, B.~W., Wise, J.~H., Xu, H., \& Norman, M.~L. 2015, \apj, 807, L12

\bibitem[{Ostriker \& McKee(1988)}]{OstrikerMcKee1988}
Ostriker, J.~P. \& McKee, C.~F. 1988, Reviews of Modern Physics, 60, 1

\bibitem[{Pakmor {et~al.}(2016)Pakmor, Springel, Bauer, Mocz, Munoz, Ohlmann,
  Schaal, \& Zhu}]{Pakmor+2016}
Pakmor, R., Springel, V., Bauer, A., {et~al.} 2016, \mnras, 455, 1134

\bibitem[{{Planck Collaboration} {et~al.}(2016){Planck Collaboration}, Ade,
  Aghanim, Arnaud, Ashdown, Aumont, Baccigalupi, Banday, Barreiro, Bartlett,
  Bartolo, Battaner, Battye, Benabed, Beno{\^i}t, {Benoit-L{\'e}vy}, Bernard,
  Bersanelli, Bielewicz, Bock, Bonaldi, Bonavera, Bond, Borrill, Bouchet,
  Boulanger, Bucher, Burigana, Butler, Calabrese, Cardoso, Catalano, Challinor,
  Chamballu, Chary, Chiang, Chluba, Christensen, Church, Clements, Colombi,
  Colombo, Combet, Coulais, Crill, Curto, Cuttaia, Danese, Davies, Davis, {de
  Bernardis}, {de Rosa}, {de Zotti}, Delabrouille, D{\'e}sert, Di~Valentino,
  Dickinson, Diego, Dolag, Dole, Donzelli, Dor{\'e}, Douspis, Ducout, Dunkley,
  Dupac, Efstathiou, Elsner, En{\ss}lin, Eriksen, Farhang, Fergusson, Finelli,
  Forni, Frailis, Fraisse, Franceschi, Frejsel, Galeotta, Galli, Ganga,
  Gauthier, Gerbino, Ghosh, Giard, {Giraud-H{\'e}raud}, Giusarma, Gjerl{\o}w,
  {Gonz{\'a}lez-Nuevo}, G{\'o}rski, Gratton, Gregorio, Gruppuso, Gudmundsson,
  Hamann, Hansen, Hanson, Harrison, Helou, {Henrot-Versill{\'e}},
  {Hern{\'a}ndez-Monteagudo}, Herranz, Hildebrandt, Hivon, Hobson, Holmes,
  Hornstrup, Hovest, Huang, Huffenberger, Hurier, Jaffe, Jaffe, Jones, Juvela,
  Keih{\"a}nen, Keskitalo, Kisner, Kneissl, Knoche, Knox, Kunz, {Kurki-Suonio},
  Lagache, L{\"a}hteenm{\"a}ki, Lamarre, Lasenby, Lattanzi, Lawrence, Leahy,
  Leonardi, Lesgourgues, Levrier, Lewis, Liguori, Lilje, {Linden-V{\o}rnle},
  {L{\'o}pez-Caniego}, Lubin, {Mac{\'i}as-P{\'e}rez}, Maggio, Maino, Mandolesi,
  Mangilli, Marchini, Maris, Martin, Martinelli, {Mart{\'i}nez-Gonz{\'a}lez},
  Masi, Matarrese, McGehee, Meinhold, Melchiorri, Melin, Mendes, Mennella,
  Migliaccio, Millea, Mitra, {Miville-Desch{\^e}nes}, Moneti, Montier,
  Morgante, Mortlock, Moss, Munshi, Murphy, Naselsky, Nati, Natoli,
  Netterfield, {N{\o}rgaard-Nielsen}, Noviello, Novikov, Novikov, Oxborrow,
  Paci, Pagano, Pajot, Paladini, Paoletti, Partridge, Pasian, Patanchon,
  Pearson, Perdereau, Perotto, Perrotta, Pettorino, Piacentini, Piat,
  Pierpaoli, Pietrobon, Plaszczynski, Pointecouteau, Polenta, Popa, Pratt,
  Pr{\'e}zeau, Prunet, Puget, Rachen, Reach, Rebolo, Reinecke, Remazeilles,
  Renault, Renzi, Ristorcelli, Rocha, Rosset, Rossetti, Roudier, {Rouill{\'e}
  d'Orfeuil}, {Rowan-Robinson}, {Rubi{\~n}o-Mart{\'i}n}, Rusholme, Said,
  Salvatelli, Salvati, Sandri, Santos, Savelainen, Savini, Scott, Seiffert,
  Serra, Shellard, Spencer, Spinelli, Stolyarov, Stompor, Sudiwala, Sunyaev,
  Sutton, {Suur-Uski}, Sygnet, Tauber, Terenzi, Toffolatti, Tomasi, Tristram,
  Trombetti, Tucci, Tuovinen, T{\"u}rler, Umana, Valenziano, Valiviita,
  Van~Tent, Vielva, Villa, Wade, Wandelt, Wehus, White, White, Wilkinson, Yvon,
  Zacchei, \& Zonca}]{PlanckCollaboration+2016}
{Planck Collaboration}, Ade, P. A.~R., Aghanim, N., {et~al.} 2016, \aap, 594,
  A13

\bibitem[{Ritter {et~al.}(2016)Ritter, {Safranek-Shrader}, Milosavljevi{\'c},
  \& Bromm}]{Ritter+2016}
Ritter, J.~S., {Safranek-Shrader}, C., Milosavljevi{\'c}, M., \& Bromm, V.
  2016, \mnras, 463, 3354

\bibitem[{Ritter {et~al.}(2015)Ritter, Sluder, {Safranek-Shrader},
  Milosavljevi{\'c}, \& Bromm}]{Ritter+2015}
Ritter, J.~S., Sluder, A., {Safranek-Shrader}, C., Milosavljevi{\'c}, M., \&
  Bromm, V. 2015, \mnras, 451, 1190

\bibitem[{Saccardi {et~al.}(2023)Saccardi, Salvadori, D'Odorico, Cupani,
  Fumagalli, Berg, Becker, Ellison, \& Lopez}]{Saccardi+2023}
Saccardi, A., Salvadori, S., D'Odorico, V., {et~al.} 2023, \apj, 948, 35

\bibitem[{Salvadori {et~al.}(2019)Salvadori, Bonifacio, Caffau, Korotin,
  Andreevsky, Spite, \& Sk{\'u}lad{\'o}ttir}]{Salvadori+2019}
Salvadori, S., Bonifacio, P., Caffau, E., {et~al.} 2019, \mnras, 487, 4261

\bibitem[{Salvadori {et~al.}(2023)Salvadori, D'Odorico, Saccardi, Skuladottir,
  \& Vanni}]{Salvadori+2023}
Salvadori, S., D'Odorico, V., Saccardi, A., Skuladottir, A., \& Vanni, I. 2023,
  \memsai, 94, 215

\bibitem[{Sasaki {et~al.}(2014)Sasaki, Clark, Springel, Klessen, \&
  Glover}]{Sasaki+2014}
Sasaki, M., Clark, P.~C., Springel, V., Klessen, R.~S., \& Glover, S. C.~O.
  2014, \mnras, 442, 1942

\bibitem[{Schaerer(2002)}]{Schaerer2002}
Schaerer, D. 2002, \aap, 382, 28

\bibitem[{Schauer {et~al.}(2019)Schauer, Glover, Klessen, \&
  Ceverino}]{Schauer+2019}
Schauer, A. T.~P., Glover, S. C.~O., Klessen, R.~S., \& Ceverino, D. 2019,
  \mnras, 484, 3510

\bibitem[{Schauer {et~al.}(2021)Schauer, Glover, Klessen, \&
  Clark}]{Schauer+2021}
Schauer, A. T.~P., Glover, S. C.~O., Klessen, R.~S., \& Clark, P. 2021, \mnras,
  507, 1775

\bibitem[{Sharda \& Menon(2025)}]{ShardaMenon2025}
Sharda, P. \& Menon, S.~H. 2025, \mnras, 540, 1745

\bibitem[{Sharda {et~al.}(2025)Sharda, Menon, Gerasimov, Bromm, Burkhart,
  Haemmerl{\'e}, {van Veenen}, \& Wibking}]{Sharda+2025}
Sharda, P., Menon, S.~H., Gerasimov, R., {et~al.} 2025, \mnras, 541, L1

\bibitem[{Sk{\'u}lad{\'o}ttir {et~al.}(2024)Sk{\'u}lad{\'o}ttir, Koutsouridou,
  Vanni, Amarsi, Lucchesi, Salvadori, \& Aguado}]{Skuladottir+2024a}
Sk{\'u}lad{\'o}ttir, {\'A}., Koutsouridou, I., Vanni, I., {et~al.} 2024, \apj,
  968, L23

\bibitem[{Springel(2010)}]{Springel2010}
Springel, V. 2010, \mnras, 401, 791

\bibitem[{Stacy \& Bromm(2013)}]{StacyBromm2013}
Stacy, A. \& Bromm, V. 2013, \mnras, 433, 1094

\bibitem[{Str{\"o}mgren(1939)}]{Stromgren1939}
Str{\"o}mgren, B. 1939, \apj, 89, 526

\bibitem[{Suda {et~al.}(2017)Suda, Hidaka, Aoki, Katsuta, Yamada, Fujimoto,
  Ohtani, Masuyama, Noda, \& Wada}]{Suda+2017}
Suda, T., Hidaka, J., Aoki, W., {et~al.} 2017, \pasj, 69, 76

\bibitem[{Takahashi {et~al.}(2018)Takahashi, Yoshida, \&
  Umeda}]{Takahashi+2018}
Takahashi, K., Yoshida, T., \& Umeda, H. 2018, \apj, 857, 111

\bibitem[{Tarumi {et~al.}(2020)Tarumi, Hartwig, \& Magg}]{Tarumi+2020}
Tarumi, Y., Hartwig, T., \& Magg, M. 2020, \apj, 897, 58

\bibitem[{Tress {et~al.}(2020)Tress, Sormani, Glover, Klessen, Battersby,
  Clark, Hatchfield, \& Smith}]{Tress+2020}
Tress, R.~G., Sormani, M.~C., Glover, S. C.~O., {et~al.} 2020, \mnras, 499,
  4455

\bibitem[{Truelove {et~al.}(1997)Truelove, Klein, McKee, Holliman, Howell, \&
  Greenough}]{Truelove+1997}
Truelove, J.~K., Klein, R.~I., McKee, C.~F., {et~al.} 1997, \apj, 489, L179

\bibitem[{Tseliakhovich \& Hirata(2010)}]{TseliakhovichHirata2010}
Tseliakhovich, D. \& Hirata, C. 2010, \prd, 82, 083520

\bibitem[{Turk {et~al.}(2011)Turk, Smith, Oishi, Skory, Skillman, Abel, \&
  Norman}]{Turk+2011}
Turk, M.~J., Smith, B.~D., Oishi, J.~S., {et~al.} 2011, \apjs, 192, 9

\bibitem[{Vanni {et~al.}(2024)Vanni, Salvadori, D'Odorico, Becker, \&
  Cupani}]{Vanni+2024}
Vanni, I., Salvadori, S., D'Odorico, V., Becker, G.~D., \& Cupani, G. 2024,
  \apj, 967, L22

\bibitem[{Vanzella {et~al.}(2023)Vanzella, Loiacono, Bergamini, Me{\v
  s}tri{\'c}, Castellano, Rosati, Meneghetti, Grillo, Calura, Mignoli, Brada{\v
  c}, Adamo, Rihtar{\v s}i{\v c}, Dickinson, Gronke, Zanella, Annibali,
  Willott, Messa, Sani, Acebron, Bolamperti, Comastri, Gilli, Caputi, Ricotti,
  Gruppioni, Ravindranath, Mercurio, Strait, Martis, Pascale, Caminha,
  Annunziatella, \& Nonino}]{Vanzella+2023}
Vanzella, E., Loiacono, F., Bergamini, P., {et~al.} 2023, \aap, 678, A173

\bibitem[{Ventura {et~al.}(2024)Ventura, Qin, Balu, \& Wyithe}]{Ventura+2024}
Ventura, E.~M., Qin, Y., Balu, S., \& Wyithe, J. S.~B. 2024, \mnras, 529, 628

\bibitem[{Virtanen {et~al.}(2020)Virtanen, Gommers, Oliphant, Haberland, Reddy,
  Cournapeau, Burovski, Peterson, Weckesser, Bright, {van der Walt}, Brett,
  Wilson, Millman, Mayorov, Nelson, Jones, Kern, Larson, Carey, Polat, Feng,
  Moore, VanderPlas, Laxalde, Perktold, Cimrman, Henriksen, Quintero, Harris,
  Archibald, Ribeiro, Pedregosa, {van Mulbregt}, \& {SciPy 1. 0
  Contributors}}]{Virtanen+2020}
Virtanen, P., Gommers, R., Oliphant, T.~E., {et~al.} 2020, Nature Methods, 17,
  261

\bibitem[{Weinberger {et~al.}(2020)Weinberger, Springel, \&
  Pakmor}]{Weinberger+2020}
Weinberger, R., Springel, V., \& Pakmor, R. 2020, \apjs, 248, 32

\bibitem[{Whalen {et~al.}(2013)Whalen, Even, Frey, Smidt, Johnson, Lovekin,
  Fryer, Stiavelli, Holz, Heger, Woosley, \& Hungerford}]{Whalen+2013}
Whalen, D.~J., Even, W., Frey, L.~H., {et~al.} 2013, \apj, 777, 110

\bibitem[{Woosley(2017)}]{Woosley2017}
Woosley, S.~E. 2017, \apj, 836, 244

\bibitem[{Xing {et~al.}(2023)Xing, Zhao, Liu, Heger, Han, Aoki, Chen, Ishigaki,
  Li, \& Zhao}]{Xing+2023}
Xing, Q.-F., Zhao, G., Liu, Z.-W., {et~al.} 2023, \nat, 618, 712

\bibitem[{Yoshida {et~al.}(2006)Yoshida, Omukai, Hernquist, \&
  Abel}]{Yoshida+2006}
Yoshida, N., Omukai, K., Hernquist, L., \& Abel, T. 2006, \apj, 652, 6

\bibitem[{Zackrisson {et~al.}(2024)Zackrisson, Hultquist, Kordt, Diego,
  Nabizadeh, Vikaeus, Meena, Zitrin, Volpato, Lundqvist, Welch, Costa, \&
  Windhorst}]{Zackrisson+2024}
Zackrisson, E., Hultquist, A., Kordt, A., {et~al.} 2024, \mnras, 533, 2727

\end{thebibliography}

\appendix
\section{Consistency with the Complementary Simulation}
\begin{figure}
	\centering
	\includegraphics[width=\linewidth]{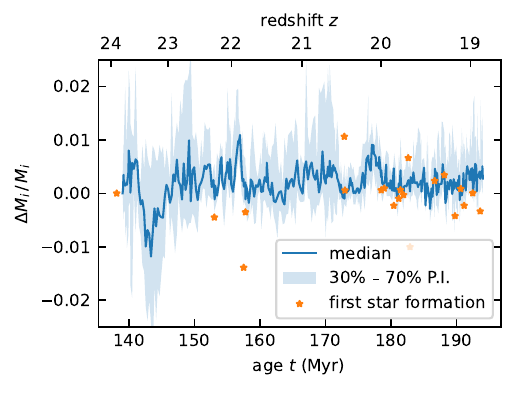}
	\caption{The quantity $\Delta M_i = M_{i,\textrm{this work}} - M_{i,\textrm{Magg+2022}}$ compares the halo masses in this simulation with the results presented in \citet{Magg+2022}.
		The median and the 30th to 70th percentile scatter by \qty{2}{\percent} or less around 0.
		Additionally, the halo mass differences at the time when the first star forms are highlighted, too.}
	\label{fig:massconsistency}
\end{figure}

This simulation is complementary to that presented by \citet{Magg+2022} and therefore we expect the global evolution and times of first star formation to be largely similar.
We reprocess the data by \citet{Magg+2022} with ROCKSTAR and CONSISTENT TREES adopting identical parameters.
We compute the mass difference $\Delta M_i = M_{i,\textrm{this work}} - M_{i,\textrm{Magg+2022}}$ of the dark matter virial masses $M_i$ where $i$ indicates different star-forming halos.
The snapshot cadence in \citet{Magg+2022} is smaller than in this work, and the mass evolution of the main progenitor branch in \citet{Magg+2022} is linearly interpolated at the snapshot times of this work.
In Fig.~\ref{fig:massconsistency}, we compare the masses of the dark matter minihalos $\Delta M_i/M_{i,\textrm{this work}}$ by showing the median and 30th to 70th percentile.
The consistency of the dark matter minihalo masses is within \qtyrange{1}{2}{\percent} and at the times of star formation, the maximum difference is \qty{1.4}{\percent}.

For each star-forming halo, we have also computed the difference in the location of the halo centre of mass. This has a mean value of approximately \qty{0.18}{\codelength} with a standard deviation of \qty{0.37}{\codelength} which is likely due to minor differences in the distribution of matter within the virial radius. For comparison, the virial radii of these halos are \qtyrange{2}{4}{\codelength}, i.~e., the differences are small. 

\section{Metal enrichment evolution}
\begin{figure}
	\centering
	\includegraphics[width=\linewidth]{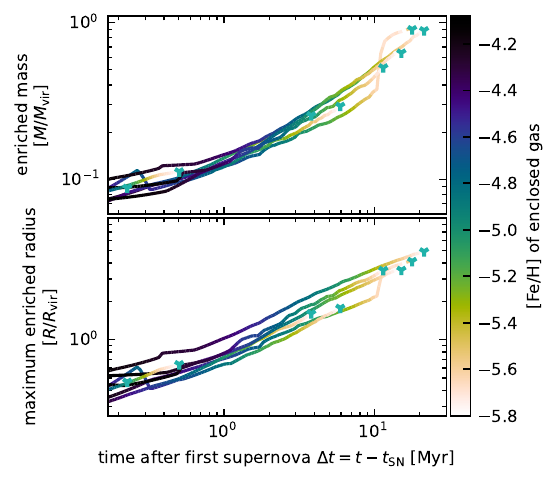}
	\caption{Same as Fig.~\ref{fig:enrichment} but normalised to the virial mass and virial radius.}
	\label{fig:enrichmentnorm}
\end{figure}

\begin{figure}
	\centering
	\includegraphics[width=\linewidth]{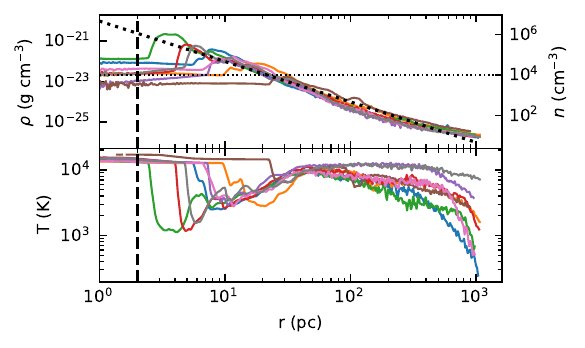}
	\caption{
		Radial profiles of the recollapsing halos at the time of the supernova explosion.
		The upper panel shows the density and the lower panel the temperature in the radial shells.
		The exponential fit $\log \rho \propto -k_\rho \log r$ to the density profile outside the Strömgren sphere in the upper panel has the exponent $k_\rho=2.0$.
	}
	\label{fig:radialprofshells}
\end{figure}

We show the evolution of an upper limit of the enriched mass and the maximum enriched radius normalised to the virial mass and the virial radius, respectively, in Fig.~\ref{fig:enrichmentnorm}.
The gas density and temperature profile of the halos just before the supernova explosion is shown in Fig.~\ref{fig:radialprofshells}. 
The fit to the density profile is consistent with $\rho\propto r^{-2}$, and it is used to determine the expansion of the supernova blast wave in the Sedov-Taylor theory.
\end{document}